\documentclass[12pt]{article}
\pdfoutput=1
\usepackage{putex}
\usepackage{amssymb}
\usepackage{amsmath}
\usepackage{comment}
\usepackage{empheq}
\usepackage{epsf}%\usepackage{showlabels}
\usepackage{graphicx}
\usepackage{epstopdf}
\usepackage{enumerate}
\usepackage{empheq}
\usepackage{cite}
\usepackage{caption}
\usepackage{subcaption}
\usepackage{hyperref}
\usepackage[utf8x]{inputenc}
\usepackage{braket}
\usepackage{titlesec}
\usepackage{fixltx2e}
\usepackage{cleveref}
\hypersetup{breaklinks}
\usepackage[usenames,dvipsnames,svgnames,table]{xcolor}
\usepackage{tikz}
\usetikzlibrary{decorations.markings}
\tikzset{->-/.style={decoration={
  markings,
  mark=at position .5 with {\arrow{>}}},postaction={decorate}}}

\newlength\mytemplen
\newsavebox\mytempbox

\makeatletter
\newcommand\mybluebox{%
    \@ifnextchar[%]
       {\@mybluebox}%
       {\@mybluebox[0pt]}}

\def\@mybluebox[#1]{%
    \@ifnextchar[%]
       {\@@mybluebox[#1]}%
       {\@@mybluebox[#1][0pt]}}

\def\@@mybluebox[#1][#2]#3{
    \sbox\mytempbox{#3}%
    \mytemplen\ht\mytempbox
    \advance\mytemplen #1\relax
    \ht\mytempbox\mytemplen
    \mytemplen\dp\mytempbox
    \advance\mytemplen #2\relax
    \dp\mytempbox\mytemplen
    {\hspace{1em}\usebox{\mytempbox}\hspace{1em}}}

\makeatother

   \def\Rc{{\cal R}}

\def\P{\Phi}

\def\DP{\Delta_{\Phi}}

\def\anl{a_{n,\ell}}
\def\gnl{\gamma_{n,\ell}}
\def\o{\over}

\def\d{\delta}

\def\0{{(0)}}
\def\1{{(1)}}
\def\2{{(2)}}
\def\3{{(3)}}
\def\4{{(4)}}

\def\G{\Gamma}

\def\eqr{\eqref}
\def\sec{\section}

\def\l{\lambda}

\def\a{\alpha}
\def\rar{\rightarrow}

\def\la{\langle}
\def\ra{\rangle}
\def\O{{\cal O}}

\def\Gc{{\cal G}}

\def\ssec{\subsection}
\def\sssec{\subsubsection}
\def\sec{\section}
\newcommand{\Farg}[3]{\left( \begin{array}{c} #1 \\ #2\end{array} \Big| #3 \right)}
\def\i{\infty}

\def\F{{\cal F}}
\def\O{{\cal O}}
\def\ra{\rangle}
\def\la{\langle}

\def\Db{\bar{D}}
\def\t{\tau}

\def\s{\sigma}

\def\D{\Delta}

\def\g{\gamma}
\def\a{\alpha}
\newcommand {\be} {\begin {equation}}
\newcommand {\ee} {\end {equation}}

\newcommand {\bes} {\begin {equation*}}
\newcommand {\ees} {\end {equation*}}

\newcommand{\es}[2] {\begin{equation} \label{#1} \begin{split} #2 \end{split} \end{equation}}
\newcommand{\e}[2] {\begin{equation} \label{#1} #2 \end{equation}}

\newcommand{\Z}{\mathbb{Z}}
\newcommand{\N}{\mathbb{N}}
\newcommand{\R}{\mathbb{R}}
\newcommand{\C}{\mathbb{C}}

\newcommand{\beq}{\begin{equation}}
\newcommand{\eeq}{\end{equation}}

\newcommand{\cF}{\mathcal{F}}

\newcommand{\p}{\partial}

\def\be{ \begin{equation} }
\def\ee{ \end{equation} }
\def\N{{\cal N}}

\def\C{{\cal C}}
\def\cF{{\cal F}}

\def\eqr{\eqref}

\def\b{\beta}
\def\l{\lambda}
\def\C{{\cal C}}

\newcommand\ov{\over}

\newcommand\zb{\bar{z}}

\def\rar{\rightarrow}
\def\C{\mathcal{C}}

\def\zb{\overline{z}}

\def\eps{\epsilon}

 \newmuskip\pFqmuskip

\newcommand*\pFq[6][8]{%
  \begingroup % only local assignments
  \pFqmuskip=#1mu\relax
  \mathcode`\,=\string"8000
  \begingroup\lccode`\~=`\,
  \lowercase{\endgroup\let~}\pFqcomma
  {}_{#2}F_{#3}{\left[\genfrac..{0pt}{}{#4}{#5};#6\right]}%
  \endgroup
}
\newcommand{\pFqcomma}{\mskip\pFqmuskip}
\makeatletter
\renewcommand{\@maketitle}{
\newpage
 \begin{center}%
  {\large\bfseries \@title \par}%
 \end{center}%
 \par} \makeatother
\numberwithin{equation}{section}

\begin{document}

\preprint{PUPT-2548\\CALT-TH 2018-001}

\institution{PU}{Department of Physics, Princeton University, Princeton, NJ 08544, USA}
\institution{CT}{Walter Burke Institute for Theoretical Physics, California Institute of Technology, \cr Pasadena, CA, 91125, USA}
\institution{ITEP}{ITEP, B. Cheremushkinskaya 25, Moscow, 117218, Russia}

\title{Double-Trace Deformations of\\ Conformal Correlations}

\authors{Simone Giombi\worksat{\PU}, Vladimir Kirilin\worksat{\PU,\ITEP}, Eric Perlmutter\worksat{\PU, \CT}}

\abstract{Large $N$ conformal field theories often admit unitary renormalization group flows triggered by double-trace deformations. We compute the change in scalar four-point functions under double-trace flow, to leading order in $1/N$. This has a simple dual in AdS, where the flow is implemented by a change of boundary conditions, and provides a physical interpretation of single-valued conformal partial waves. We extract the change in the conformal dimensions and three-point coefficients of infinite families of double-trace composite operators. Some of these quantities are found to be sign-definite under double-trace flow. As an application, we derive anomalous dimensions of spinning double-trace operators comprised of non-singlet constituents in the $O(N)$ vector model.}

\date{ }

\maketitle
\setcounter{tocdepth}{2}
\tableofcontents

\sec{Introduction and Summary}

In quantum field theory, there are few instances where, absent extra symmetries or dualities, one can analytically compute changes in observables under renormalization group (RG) flow between conformal fixed points. In the presence of a small parameter, more progress is possible. A special class of large $N$ RG flows in this category are those triggered by a ``double-trace'' deformation of a conformal field theory (CFT). These flows have long been part of the vector model paradigm, in which the critical boson and fermion theories \cite{Wilson:1971dc, Gross:1974jv} may be reached by double-trace deformations of free bosons and fermions, respectively. They also have a natural interpretation in the context of AdS/CFT, where they are generated by a change in boundary conditions \cite{Klebanov:1999tb, wittendt, Klebanov:2002ja}. 

The effect of such deformations at the level of the CFT partition function \cite{Gubser:2002vv}, and the relation to AdS boundary conditions \cite{Klebanov:1999tb, wittendt, Berkooz:2002ug, Mueck:2002gm, Gubser:2002zh, Gubser:2002vv, Hartman:2006dy, Diaz:2007an}, was clearly laid out in the early days of AdS/CFT. In this work, inspired by a  contemporary perspective on the value of correlation functions to the foundations of CFT, we compute some more complicated quantities: the change in CFT four-point functions under double-trace flow from UV to IR, and the effect of such flows on the sector of double-trace operators in the IR CFT. The calculations have simple bulk duals in AdS, an elegant relation to harmonic functions for the conformal group, and reveal new observables whose change from UV to IR is sign-definite.

\begin{figure}
    \centering
       \includegraphics[width = .6\textwidth]{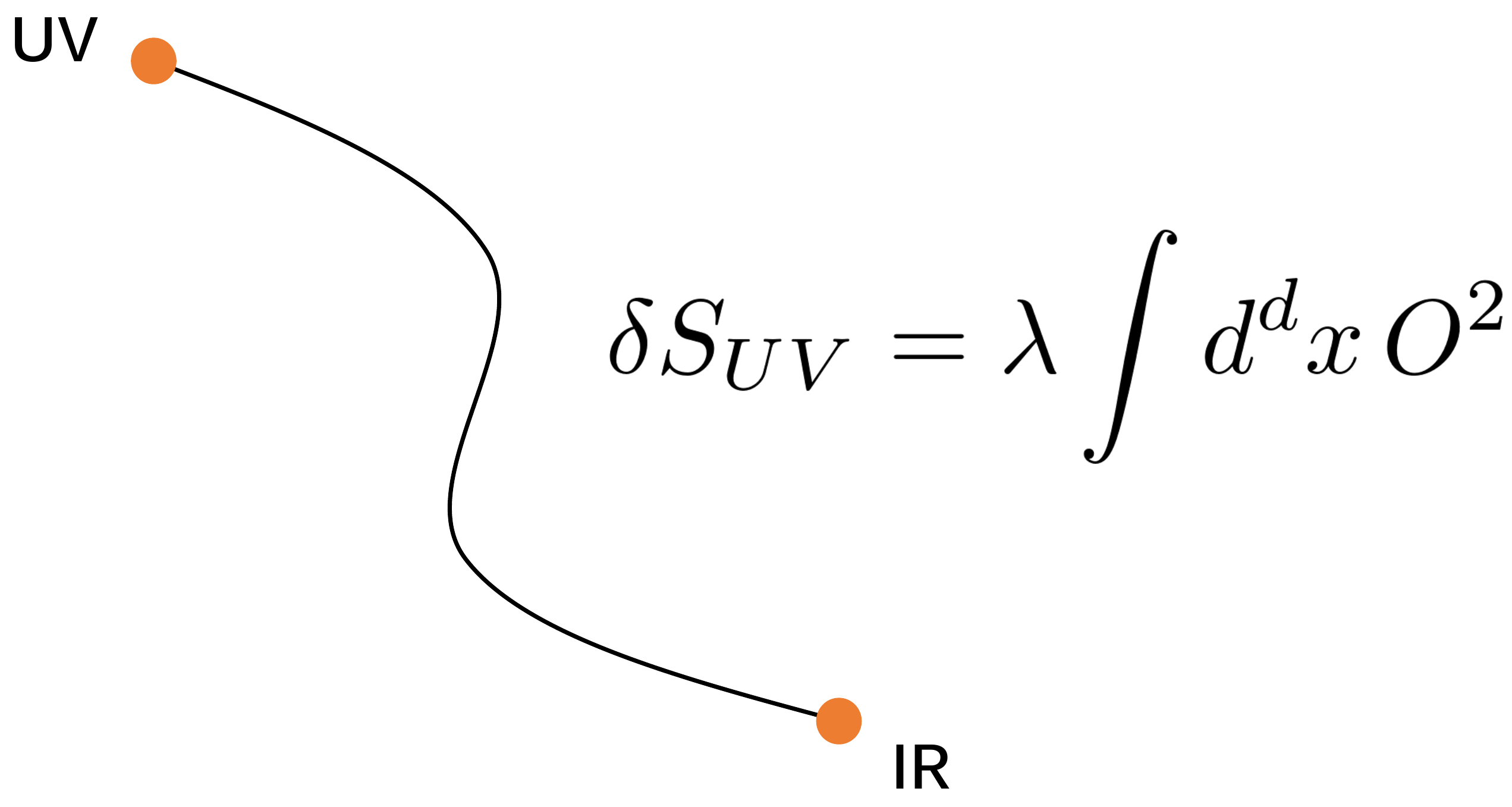}
         \caption{A renormalization group flow triggered by a double-trace deformation of a large-$N$ CFT: the UV fixed point, at which the operator $O$ has $\D<d/2$, flows to an IR fixed point, at which $O$ has conformal dimension $d-\D$, at leading order in $1/N$. In this paper, we compute the leading order change in four-point functions of single-trace operators, other than $O$, that couple to $O$.}\label{fig1}
\end{figure}

A double-trace flow is defined as follows. Consider a CFT$_d$ that admits a large $N$ expansion, and possesses a single-trace scalar operator $O $ with $\Delta<d/2$. Then we can consider the relevant double-trace deformation 
\begin{equation}\label{eq12}
S_{\lambda} = S_{CFT}+\lambda \int d^d x \,O ^2
\end{equation}
This triggers a flow from the unperturbed CFT in the UV to a new CFT in the IR in which the operator $O $ has dimension $d-\Delta+\O(1/N)$. (We use the ``vector model convention'' $C_T\sim N$ throughout this work.) See Figure \ref{fig1}. 
On the AdS side, this has a clean interpretation: it corresponds to changing the boundary conditions on the bulk scalar field $\varphi$ dual to $O$ \cite{Klebanov:1999tb, Klebanov:1998hh}. The bulk scalar has mass given by $m^2L_{\rm AdS}^2=\Delta(\Delta-d)$, and when $d/2-1<\Delta<d/2$ it admits two unitary boundary conditions: the choice $\varphi \sim z^{\Delta}$ (with the AdS boundary in Poincar\'e coordinates at $z=0$) corresponds to the UV CFT, while $\varphi \sim z^{d-\Delta}$ corresponds to the IR CFT reached after the double-trace perturbation.   

Double-trace flows are special: in the $1/N$ expansion (where the notion of ``double-trace'' is well-defined), the IR CFT is only mildly different than its UV counterpart. To leading order in $1/N$, the single-trace dimensions and OPE coefficients are identical except for those involving $O$. Beyond leading order, all operator data is generically modified. This includes the dimensions and OPE coefficients of double-trace composite operators {\it not} comprised of $O$. Given some other single-trace operator $\P$, there exist infinite towers of double-trace primary operators of schematic form
\e{dtprim}{[\Phi\Phi]_{n,\ell} \approx\,\,: \!\Phi \p^{2n}\p_{\mu_1}\ldots \p_{\mu_\ell}\Phi\!:-~(\text{traces})~.}
where the subtraction enforces the primary condition. There is one operator for each $(n,\ell)$, with conformal dimensions
\e{}{\D_{n,\ell} = 2\DP + 2n+\ell+\g_{n,\ell}}
for some $\g_{n,\ell}$. We now know that their quantum numbers are a rich source of dynamical information about interacting large $N$ CFTs: they are sensitive to the coupling strength; they are constrained by causality and unitarity \cite{Fitzpatrick:2010zm, Hartman:2015lfa, Komargodski:2016gci}; crossing symmetry of four-point functions relates the exchange of double-trace operators to Regge trajectories of single-trace operators \cite{Li:2017lmh,Alday:2015ota,Kulaxizi:2017ixa}, and hence $\g_{n,\ell}$ to the cusp anomalous dimension and the higher spin gap scale; and in the AdS/CFT context, the nature of inter-particle forces in the bulk is manifest in the behavior of $\g_{n,\ell}$ as a function of $(n,\ell)$, thus making $\g_{n,\ell}$ a sensitive probe of bulk locality \cite{Heemskerk:2009pn, Komargodski:2012ek,Fitzpatrick:2012yx, Maldacena:2015iua, Alday:2015ewa, Li:2015rfa, Li:2015itl, Alday:2017gde}. These reasons motivate the study of how the $\g_{n,\ell}$ change under double-trace flow. The $\g_{n,\ell}$ are $1/N$-suppressed, so it would seem more difficult to compute their change under double-trace flow. The same is true for three-point functions $\la \P\P[\P\P]_{n,\ell}\ra$. However, the change of the $\g_{n,\ell}$ from UV to IR can be extracted from the leading-order change in the connected part of the four-point funtion $\la \P\P\P\P\ra$, which itself may be straightforwardly computed.\footnote{Note that we do not compute the change from UV to IR 
of the dimension $\DP$ of the single-trace constituents. This is encoded in the $1/N$ correction to the two-point function
of $\P$, which corresponds to a one-loop diagram from the AdS point of view.}  

In Section \ref{sec2}, we begin with an exercise, in which we compute the change in three-point functions $\la OOO\ra, \la \P O O\ra$ and $\la \P\P O\ra$. These results are known -- for instance, by modifying the boundary condition in AdS calculations of CFT three-point functions -- but will appear in our later analysis and serve as a useful warmup. 

In Section \ref{s3}, we compute the change in connected four-point functions $\la \P\P\P\P\ra$ and $\la\P\Psi\P\Psi\ra$ from UV to IR, where $\P$ and $\Psi$ are distinct scalar primaries. Part of our message is that the result, which is manifestly crossing-symmetric, may be expressed simply in terms of a single $\Db$-function -- see \eqr{4pt-final} and \eqr{pair2}. The $\Db$-functions are themselves not elementary, but arise in many contexts in CFT at both weak and strong coupling (e.g. \cite{GonzalezRey:1998tk,Liu:1998ty,DHoker:1999kzh,Dolan:2000ut, Dolan:2001tt}). In Section \ref{AdS}, we review the manifest equivalence 
\cite{Hartman:2006dy, GiombiYin} between our CFT calculation and a tree-level AdS calculation. Moreover, the change in these four-point functions is technically essentially identical to the computation of four-point conformal partial waves for principal series representations, whose utility has recently been emphasized \cite{Hogervorst:2017sfd, Hogervorst:2017kbj, Gadde:2017sjg, Caron-Huot:2017vep, Murugan:2017eto, Simmons-Duffin:2017nub}.\footnote{See also e.g. 
\cite{Penedones:2010ue, Costa:2012cb, Costa:2014kfa, Bekaert:2014cea, Dyer:2017zef, Sleight:2017fpc} for 
further related work on harmonic analysis in AdS/CFT.} Thus, our result may be viewed as a novel physical interpretation of these objects in terms of double-trace flows, and provides for them a mathematical expression in terms of a $\Db$-function. We also point out that the special cases $\D+\DP=\D_\Psi-2p$ and $\D+\D_\Psi=\DP-2p$ for $p\in\Z_{\geq 0}$ involve ``extremal'' three-point functions \cite{DHoker:1999jke}, and simplify dramatically. 

In Section \ref{s4}, we use the change in four-point functions to extract the change in OPE data of the leading-twist double-trace operators $[\P\P]_{0,\ell}$ and $[\P\Psi]_{0,\ell}$, for all $\ell$. We focus mostly on anomalous dimensions $\g_\ell\equiv \g_{0,\ell}$, although we compute some OPE coefficients as well. The results for $\d\g_\ell\equiv\g^{IR}_{\ell} - \g_{\ell}^{UV}$, in \eqr{dg0ell} and \eqr{pairdgam}, have various universal features. We wish to highlight one here: for $[\P\P]_{0,\ell}$ operators, $\d\g_{0}$ as given in \eqr{g00} is independent of $\DP$. It is also always positive for unitary values of the dimensions $(\DP,\D)$. Therefore, we arrive at the interesting conclusion that this privileged class of RG flows admits sign-definite quantities over and above the usual $a/c/F$ observables. 

In Section \ref{s5}, we specify the conformal dimensions $\D$ and $\DP$ to certain values where the results simplify, and use these results as a tool to derive anomalous dimensions of some double-trace operators in the $O(N)$ vector model, in which the constituents are {\it not} $O(N)$ singlets. This is possible because $\d\g_\ell=\g_\ell^{IR}$, since the UV theory from which the $O(N)$ model descends is free. Specifically, we derive $\la \P\P^*\P\P^*\ra$ where $\P$ and $\P^*$ are conjugate operators in the $O(N)$ model, each a scalar bilinear in the rank-two symmetric traceless representation of $O(N)$, and extract $\g_\ell$ for double-trace operators $[\P\P^*]_{0,\ell}$. The result for $\g_\ell$ in various spacetime dimensions can be found in \eqr{d36}. This is, to our knowledge, a new result. It exhibits interesting harmonic behavior in $d=4-\eps$. While our technique here -- of using a calculation of $\d\g_\ell$ {\it between} fixed points to derive $\g_\ell$ {\it at} the IR fixed point -- is essentially identical to an ordinary large $N$ calculation in the $O(N)$ model, it would be interesting to apply it to other CFTs beyond the $O(N)$ model, where it acts as a simpler alternative to standard computation of a full four-point function. A recent application can be found in \cite{Giombi:2017mxl}.
 
%   This ``trick'' --  -- may be viewed, like \cite{Rychkov:2015naa}, as an alternative to traditional perturbative computations of anomalous dimensions, and it would be interesting to apply it to CFTs beyond the $O(N)$ model. 
   
Appendices \ref{appa}--\ref{conf-pert} contain some background material, various details of calculations in the text, and a conformal perturbation theory cross-check of our result for $\g_0^{IR}$ in the $O(N)$ model in $d=4-\eps$.

\sec{Warmup: Three-point functions}\label{sec2}
To introduce some formalism, and to derive a result we will use later, let us first compute the change of some three-point coefficients. All of our calculations 
will rely on the Hubbard-Stratonovich auxiliary field method, which we recall now. 

Upon introducing the deformation \eqr{eq12}, the large $N$ expansion in the IR can be developed by introducing an auxiliary field as 
\begin{equation}
S_{IR} = S_{CFT}+ \int d^d x \sigma O 
\label{S-IR}
\end{equation}
where the quadratic term $-\sigma^2/(4\lambda)$ can be dropped in the IR limit. The auxiliary field $\s$ acquires the following induced two-point function at leading order in $1/N$ (see e.g. \cite{Fei:2014yja, Diab:2016spb} for a review and more details)
\begin{equation}
\langle \sigma(x)\sigma(y)\rangle = 
-\frac{\Gamma (\Delta ) \Gamma (d-\Delta )}{\pi^d C_{OO} \Gamma \left(\frac{d}{2}-\Delta \right) \Gamma \left(\Delta -\frac{d}{2}\right)}\frac{1}{|x-y|^{2(d-\Delta)}}
\equiv \frac{C_{\sigma}}{|x-y|^{2(d-\Delta)}}\,. 
\label{sig-prop}
\end{equation}
where $O$ is normalized as
\e{}{\la O (x)O (0) \ra = {C_{OO}\o x^{2\D}}~.}
Thus in the IR limit $\sigma$, which replaces the operator $O$, becomes a scalar primary of dimension $d-\Delta+{\cal O}(1/N)$; that is, $O$ turns into its ``shadow''. All other single-trace operators have the same dimension in the UV and IR to leading order at large $N$. Correlation functions at the IR fixed point can be computed systematically in the $1/N$ expansion using the $\sigma$ propagator (\ref{sig-prop}) and the $\sigma O$ vertex in (\ref{S-IR}).\footnote{One has to be careful not to include one-loop 
bubble corrections to $\sigma$ lines, as those are already resummed when using the effective propagator (\ref{sig-prop}).} Note that (\ref{S-IR}) can be thought as a kind 
of Legendre transform relating UV and IR correlators.

To compute the change in three-point functions, we first note that all three-point functions of single-trace operators that do not involve the perturbing operator $O(x)$ are unchanged to leading order at large $N$. This is transparent in the 
AdS picture, as the corresponding tree-level three-point Witten diagrams that do not involve $O(x)$ are unaffected by the change in boundary conditions. 

For what follows, we will define norm-invariant squared OPE coefficients
\e{a123}{a_{\O_1\O_2\O_3} =\frac{C_{\O_1\O_2\O_3}^2}{C_{\O_1\O_1}C_{\O_2\O_2}C_{\O_3\O_3}}}

Let us first consider the OPE coefficient $C_{\Phi\Phi O}$, where $\Phi$ is a single-trace scalar operator other than $O$. Conformal three-point functions take the form
\begin{equation}
\langle \Phi(x_1)\Phi(x_2)O(x_3)\rangle_{UV} = \frac{C_{\Phi \Phi O}}{x_{12}^{2\Delta_{\Phi}-\Delta} x_{23}^{\Delta}x_{31}^{\Delta}}
\label{3pt-UV}
\end{equation}
In the IR we replace $O$ by $\sigma$, giving the three-point function
\begin{equation}
\begin{aligned}
\langle \Phi(x_1)\Phi(x_2)\sigma(x_3)\rangle_{IR} &= 
-\int d^d z \frac{C_{\sigma}}{|x_3-z|^{2(d-\Delta)}} \langle \Phi(x_1)\Phi(x_2)O(z)\rangle_{UV}+\ldots \\
&=-\int d^d z \frac{C_{\sigma}}{|x_3-z|^{2(d-\Delta)}}
\frac{C_{\Phi \Phi O}}{x_{12}^{2\Delta_{\Phi}-\Delta}(x_1-z)^{\Delta}(x_2-z)^{\Delta}}+\ldots 
\end{aligned}
\end{equation}
where $\ldots$ denotes subleading orders in $1/N$.  
 \begin{figure}
    \centering
       \includegraphics[width = .35\textwidth]{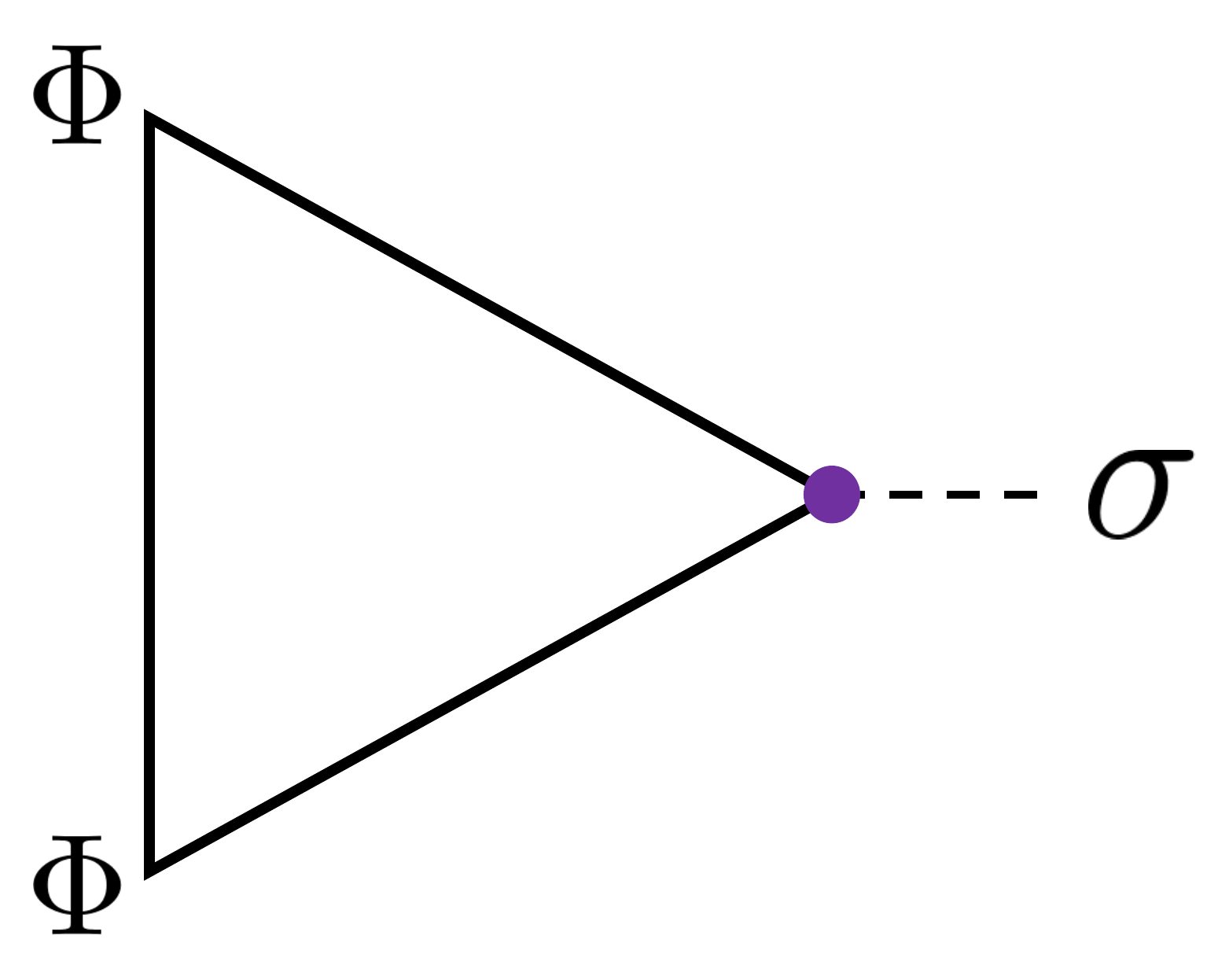}
         \caption{The triangle diagram determines the three-point coupling $\la \P\P\sigma\ra$, to which the UV coupling $\la \P\P O\ra$ flows. The purple point is integrated over.}\label{fig2}
\end{figure}
To leading order at large $N$, we evaluate this expression using the three-point conformal integral (e.g. \cite{Dolan:2000uw,Dolan:2000ut}),
\begin{equation}
\begin{aligned}\label{3ptint}
%I_{\Delta_1\Delta_2\Delta_3}(x_1,x_2,x_3)&\equiv 
\int d^d z\frac{1}{(x_1-z)^{2\Delta_1}(x_2-z)^{2\Delta_2}(x_3-z)^{2\Delta_3}}~\stackrel{\sum \Delta_i = d}{=}~
\frac{\pi^{\frac{d}{2}}a(\Delta_1)a(\Delta_2)a(\Delta_3)}{x_{12}^{d-2\Delta_3}x_{23}^{d-2\Delta_1}x_{31}^{d-2\Delta_2} }\,,
\end{aligned}
\end{equation}
where
\e{}{a(\Delta_i)\equiv\frac{\Gamma(d/2-\Delta_i)}{\Gamma(\Delta_i)}}
Specialized to our case, we thus obtain
\begin{equation}
\langle \Phi(x_1)\Phi(x_2)\sigma(x_3)\rangle_{IR} = 
\left(-C_{\sigma}C_{\Phi\Phi O}
\frac{\pi ^{d/2} \Gamma^2 \left(\frac{d-\Delta }{2}\right) \Gamma \left(\Delta -\frac{d}{2}\right)}{\Gamma^2 \left(\frac{\Delta }{2}\right) \Gamma (d-\Delta )}\right)
\frac{1}{x_{12}^{2\Delta_{\Phi}-d+\Delta}x_{23}^{d-\Delta}x_{31}^{d-\Delta}} 
%\equiv \frac{C_{\Phi\Phi\sigma}}{x_{12}^{2\Delta_{\Phi}-d+\Delta}x_{23}^{d-\Delta}x_{31}^{d-\Delta}}
\end{equation}
The factor in parenthesis is the OPE coefficient $C_{\P\P\sigma}$. Using the normalized squared OPE coefficients introduced in \eqr{a123},
\begin{equation}\label{auvppo}
a_{\Phi\Phi O}^{UV} =\frac{C_{\Phi\Phi O}^2}{C_{\Phi\Phi}^2 C_{OO}}\,,\qquad 
a_{\Phi\Phi O}^{IR} =\frac{C_{\Phi\Phi \sigma}^2}{C_{\Phi\Phi}^2 C_{\sigma}~,}
\end{equation}
the above calculation gives 
\begin{equation}
a_{\Phi\Phi O}^{IR} =- \frac{\Gamma^4 \left(\frac{d-\Delta }{2}\right) \Gamma \left(\Delta-\frac{d}{2} \right)\Gamma (\Delta )}{\Gamma^4 \left(\frac{\Delta }{2}\right) \Gamma \left(\frac{d}{2}-\Delta \right) \Gamma (d-\Delta )} a_{\Phi\Phi O}^{UV}
\label{aUV-aIR}
\end{equation}
This can be seen to match a previous derivation using AdS integrals \cite{freedman}. Notice that $\D \leftrightarrow d - \D$ properly swaps the labels UV and IR. 

Similarly, we can compute the change from UV to IR of the $C_{\Phi OO}$ OPE coefficient by attaching two $\sigma$ lines to the UV three-point function $\langle \P O O\rangle$. This was worked out explicitly in \cite{Giombi:2017mxl}, following similar steps as described above. In terms of normalized squared OPE coefficients, the result is
\begin{equation}
a_{\Phi O O}^{IR} = \frac{\Gamma^2 (\Delta ) \Gamma^2 \left(\Delta -\frac{d}{2}\right) \Gamma ^2\left(d-\Delta -\frac{\Delta _{\Phi }}{2}\right) 
\Gamma ^2\left(\frac{d}{2}-\Delta +\frac{\Delta_{\Phi}}{2}\right)}{\Gamma^2 \left(\frac{d}{2}-\Delta \right) \Gamma^2 (d-\Delta ) \Gamma^2 \left(\Delta -\frac{\Delta _{\Phi }}{2}\right) \Gamma^2 \left(-\frac{d}{2}+\Delta +\frac{\Delta _{\Phi }}{2}\right)} a_{\Phi O O}^{UV}\,.
\end{equation} 

Finally, the three-point function $\langle OOO\rangle$ in the IR can be computed by attaching three $\sigma$ lines to the UV three-point function and using repeatedly the conformal integral (\ref{3ptint}). This yields
\begin{equation}
a^{IR}_{O O O} = -\frac{\Gamma^3 (\Delta ) \Gamma^2 \left(d-\frac{3 \Delta }{2}\right) \Gamma^6 \left(\frac{d-\Delta }{2}\right) \Gamma^3 \left(\Delta -\frac{d}{2}\right)}{\Gamma^6 \left(\frac{\Delta }{2}\right) \Gamma^3 \left(\frac{d}{2}-\Delta \right) \Gamma^3 (d-\Delta ) \Gamma^2 \left(\frac{3 \Delta }{2}-\frac{d}{2}\right)} a^{UV}_{O O O}
\end{equation}
Note that for $d=3$ and $\Delta=1$, $a^{IR}_{O O O}$ vanishes, which is a well-known result for the critical $O(N)$ model in $d=3$ \cite{Petkou:1994ad}.  

%%%%%%%%%%%%%

\section{Four-point functions from UV to IR}\label{s3}
For the following calculations, we suppose the spectrum of the UV CFT includes other single-trace scalar operators $\Phi$ and $\Psi$ with UV dimensions $\Delta_{\Phi}$ and $\D_\Psi$, respectively. We will compute the change in the connected four-point functions involving $\Phi$ and $\Psi$ under the renormalization group flow triggered by $\d S_{CFT} = \int d^d x \,O ^2$, with $\D<d/2$. In what follows, we label the difference between UV and IR observables $X$ as
\e{}{ \d X\equiv X_{IR} - X_{UV}}

\subsection{Identical operators}
For the four-point function $\la \P\P\P\P\ra$, conformal symmetry constrains this difference to take the form
\e{4ptform}{\langle \Phi(x_1)\Phi(x_2)\Phi(x_3)\Phi(x_4) \rangle_{IR} - \langle \Phi(x_1)\Phi(x_2)\Phi(x_3)\Phi(x_4) \rangle_{UV}
\equiv \frac{C_{\Phi\Phi}^2}{x_{12}^{2\Delta_{\Phi}} x_{34}^{2\Delta_{\Phi}}} \d\F(u,v)}
where $\d \F(u,v)\equiv \F_{IR}(u,v)-\F_{UV}(u,v)$ is a function of the conformal cross-ratios
\e{}{u=\frac{x_{12}^2x_{34}^2}{x_{13}^2 x_{24}^2}\,,\qquad v = \frac{x_{14}^2x_{23}^2}{x_{13}^2 x_{24}^2}}
that is undetermined by conformal symmetry. Our goal is to determine $\d\F$. 

The connected four-point function of $\Phi$ in the IR can be computed in the large $N$ expansion as
\begin{equation}
\begin{aligned}
&\langle \Phi (x_1)\Phi (x_2)\Phi (x_3)\Phi (x_4)\rangle_{IR}
= \langle \Phi (x_1)\Phi (x_2)\Phi (x_3)\Phi (x_4)\rangle_{UV} 
\\
&+\frac{1}{2}\int d^d z_1\int d^d z_2 \frac{C_{\sigma}}{|z_1-z_2|^{2(d-\Delta)}} 
\langle \Phi (x_1)\Phi (x_2)\Phi (x_3)\Phi (x_4) O (z_1)O (z_2)\rangle_{UV}+\ldots 
\end{aligned}
\end{equation}
To leading order at large $N$, we can factorize the six-point function in the second line as
\begin{equation}
\begin{aligned}
&\langle \Phi (x_1)\Phi (x_2)\Phi (x_3)\Phi (x_4) 
O (z_1)O (z_2)\rangle_{UV} \\
&\simeq \langle \Phi (x_1)\Phi (x_2) O (z_1)\rangle_{UV}
\langle\Phi (x_3)\Phi (x_4) O (z_2)\rangle_{UV}+{\rm perms}+\O(1/N)
\end{aligned}
\end{equation}
where the permutations account for the $t$- and $u$-channels. So the problem is essentially just to compute the ``two-triangle" diagram with a $\sigma$ field exchange, where each triangle is the three-point function 
$\langle \Phi \Phi O\rangle$ in the UV CFT. 
 \begin{figure}
    \centering
       \includegraphics[width = .65\textwidth]{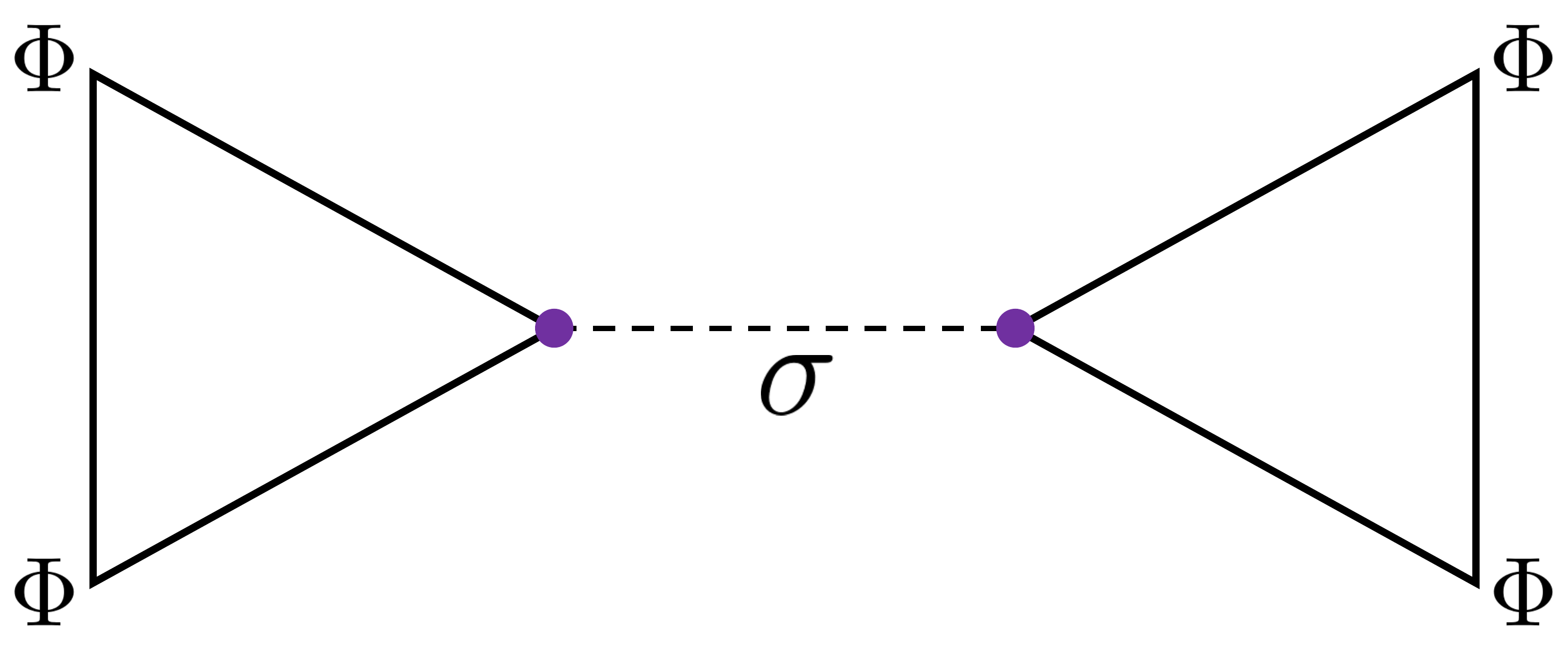}
         \caption{The two-triangle diagram, given in \eqr{2-triangle}, determines the change in the connected correlator $\la \P\P\P\P\ra$ in the $s$-channel, to leading order in $1/N$. The total result is a sum of three such diagrams, one from each channel.}\label{fig3}
\end{figure}
From the AdS point of view, this computes the difference of two four-point exchange Witten diagrams with external $\Phi$ legs and 
exchange of the bulk field dual to $O $, after taking the difference of boundary conditions on the exchanged field. We demonstrate that explicitly in Section \ref{AdS}. 

Using the form of the conformal three-point function in \eqr{3pt-UV}, the ``two-triangle" diagram is given by 
\begin{equation}
C_{\Phi \Phi O}^2 C_{\sigma} \int d^d z_1\! \int d^d z_2 \frac{1}{x_{12}^{2\Delta_{\Phi}-\Delta}|x_1-z_1|^{\Delta}|x_2-z_1|^{\Delta}}
\frac{1}{|z_1-z_2|^{2(d-\Delta)}}\frac{1}{x_{34}^{2\Delta_{\Phi}-\Delta}|x_3-z_2|^{\Delta}|x_4-z_2|^{\Delta}}
\label{2-triangle}
\end{equation}
The integration can again be performed using conformal integrals. First, we integrate in $z_1$ using the three-point integral \eqr{3ptint}. Next, we integrate over $z_2$ using the four-point conformal integral \cite{Symanzik:1972wj}
\begin{equation}
\begin{aligned}
&\int d^d z\frac{1}{(x_1-z)^{2\Delta_1}(x_2-z)^{2\Delta_2}(x_3-z)^{2\Delta_3}(x_4-z)^{2\Delta_4}}\\
&=\frac{\pi^{\frac{d}{2}}}{\prod_i \Gamma(\Delta_i)}\int_0^{\infty} \prod_i d\alpha_i \alpha_i^{\Delta_i-1} 
\frac{e^{-\frac{1}{\Lambda}\sum_{i<j}\alpha_i \alpha_j x_{ij}^2}}{\Lambda^{d/2}}\,,\qquad \quad \Lambda \equiv \sum_i \alpha_i\\
&\stackrel{\sum \Delta_i = d}{=}\frac{\pi^{\frac{d}{2}}}{\Gamma(\Delta_1)\Gamma(\Delta_2)\Gamma(\Delta_3)\Gamma(\Delta_4)} 
\frac{x_{14}^{d-2\Delta_1-2\Delta_4} x_{34}^{d-2\Delta_3-2\Delta_4}}{x_{13}^{d-2\Delta_4} x_{24}^{2\Delta_2}}
\bar{D}_{\Delta_1\Delta_2\Delta_3\Delta_4}(u,v)
\end{aligned}
\end{equation}
The function $\bar{D}_{\Delta_1\Delta_2\Delta_3\Delta_4}(u,v)$ defined above is the ubiquitous $\bar D$-function that appears in 
calculations of AdS Witten diagrams, see e.g. \cite{Dolan:2000ut} and Appendix \ref{DHG}. 
Note, however, that the $\bar D$-functions appearing here are of a special type: the sum of the exponents is equal to $d$. 

After these steps, equation (\ref{2-triangle}) yields
\begin{equation}\label{singchannel}
C_{\Phi \Phi O}^2 C_{\sigma} 
\frac{\pi^d a\left(\frac{\Delta}{2}\right)^2 a(d-\Delta)}{\Gamma^2\left(\frac{d-\Delta}{2}\right)\Gamma^2\left(\frac{\Delta}{2}\right)}
\frac{1}{x_{12}^{2\Delta_{\Phi}} x_{34}^{2\Delta_{\Phi}}} 
u^{\frac{d-\Delta}{2}} \bar{D}_{\frac{d-\Delta}{2},\frac{d-\Delta}{2},\frac{\Delta}{2},\frac{\Delta}{2}}(u,v)
\end{equation}
Note that this takes the expected conformal form. Adding all terms related by exchanging the external points, we arrive at the final, crossing-symmetric result. In terms of the normalized OPE coefficient $a_{\Phi\Phi O}^{UV}$ defined in \eqr{auvppo}, and using the definition of $C_{\sigma}$ in \eqr{sig-prop}, we can write this in the form \eqr{4ptform} with
\begin{equation}\label{4pt-final}
   \addtolength{\fboxsep}{4pt}
\boxed{
\begin{aligned}
&\d\F(u,v)=-\frac{a_{\Phi\Phi O}^{UV}\Gamma (\Delta )}{\Gamma ^4\left(\frac{\Delta }{2}\right) \Gamma \left(\frac{d}{2}-\Delta \right)}\times\\
& \left[u^{\frac{d-\Delta}{2}}\bar{D}_{\frac{d-\Delta}{2},\frac{d-\Delta}{2},\frac{\Delta}{2},\frac{\Delta}{2}}(u,v)+
u^{\Delta_{\Phi}} \bar{D}_{\frac{d-\Delta}{2},\frac{\Delta}{2},\frac{d-\Delta}{2},\frac{\Delta}{2}}(u,v) 
+\left(\frac{u}{v}\right)^{\Delta_{\Phi}} v^{\frac{d-\Delta}{2}} \bar{D}_{\frac{\Delta}{2},\frac{d-\Delta}{2},\frac{d-\Delta}{2},\frac{\Delta}{2}}(u,v)
\right]
\end{aligned}}
\end{equation}
This is one of our main results. The change in the connected part of four-point functions under double-trace flow takes a very simple form, expressed as a manifestly crossing-symmetric sum of $\bar D$-functions, one from each channel. We can equivalently write this in terms of $a^{\rm IR}_{\Phi\Phi O}$ using (\ref{aUV-aIR}). 

%\sssec{$O$ exchange}

\sssec{The $O(N)$ model four-point function}
When $\Delta=d-2$ and $\Delta_{\Phi}=(d-2)/2$, our result \eqr{4pt-final} can be compared to the connected 
correlator of the elementary fields $\phi^i$ in the critical $O(N)$ model at large $N$ 
\e{}{\langle \phi^i(x_1) \phi^j(x_2) \phi^k(x_3) \phi^l(x_4)\rangle_{\rm conn} ={\F^{ijkl}(u,v)\o (x_{12}x_{34})^{2\Delta_{\phi}}}}
upon identifying $\Phi$ with $\phi^i$, and $O$ with $\sigma \sim \phi^i\phi^i$. This is because, even though $\phi^i$ is not strictly speaking a 
``single-trace" operator, the calculation of the leading large $N$ contribution to the connected correlator takes 
the same form as the two-triangle diagram considered above, with 
the role of the triangle played by the three-point functions $\langle \phi^i(x_1) \phi^j(x_2) \phi^k\phi^k(z)\rangle$ in the UV free theory. To account for the index structure we just need to slightly generalize our calculation: adding up the singlet contributions in each channel and using $\bar D$-function identities summarized in Appendix \ref{DHG}, the expression in brackets in \eqr{4pt-final} may be written
\es{}{&\F^{ijkl}(u,v)=-\frac{2\Gamma (d-2 )}{N\Gamma ^4\left(\frac{d-2}{2}\right) \Gamma \left(2-\frac{d}{2} \right)}\times\\
&u^{\frac{d-2}{2}}\Big[\d^{ik}\d^{jl}\bar{D}_{\frac{d-2}{2},\frac{d-2}{2},1,1}(u,v)+
\d^{ij}\d^{kl} \bar{D}_{1,\frac{d-2}{2},1,\frac{d-2}{2}}(u,v) 
+\d^{il}\d^{kj}\bar{D}_{1,\frac{d-2}{2},\frac{d-2}{2},1}(u,v)\Big]+\O({1/N^2})}
%\end{equation}
%
This is indeed the correct result for the connected correlator in the critical $O(N)$ model \cite{Lang:1991kp}.\footnote{See also (159) of \cite{Alday:2015ewa}, modulo the missing power of $u$.} 

\subsection{Equivalence between CFT and AdS calculations}
\label{AdS}
 \begin{figure}
    \centering
       \includegraphics[width = \textwidth]{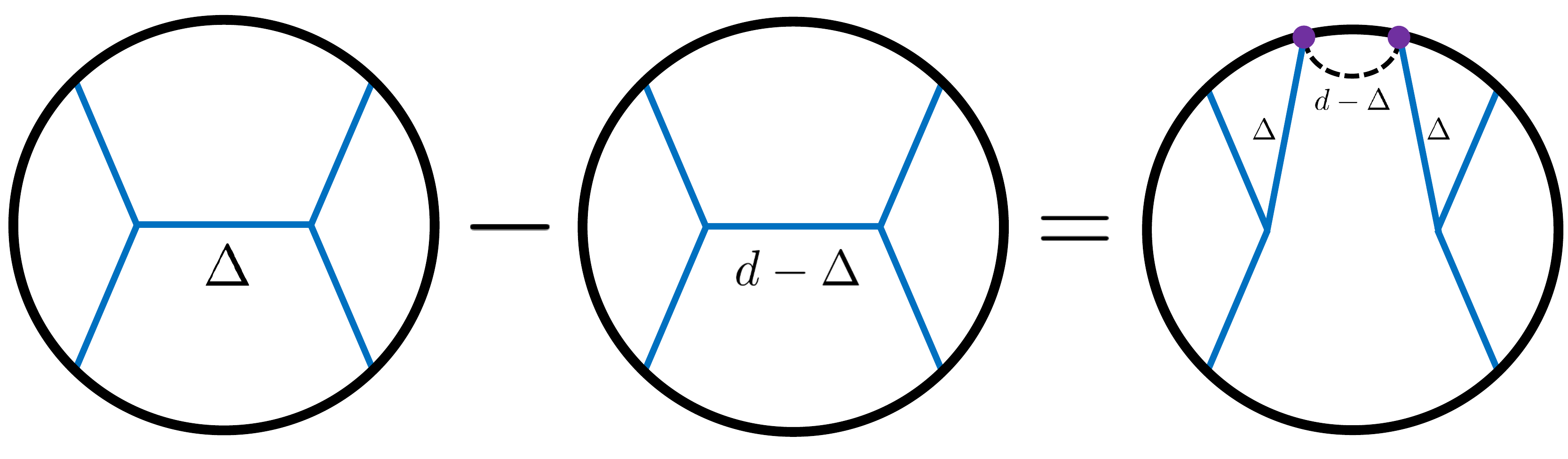}
         \caption{The AdS dual of the two-triangle diagram in CFT. The difference of two exchange diagrams with external $\P$ fields -- one with standard quantization $(\D)$ of the field dual to $O$, and one with alternate quantization $(d-\D)$ -- can be written, using the split representation of the AdS harmonic function, as a pair of boundary three-point functions tied together by a boundary two-point function of dimension $d-\D$. This is manifestly equivalent to the two-triangle diagram in CFT.}\label{fig4}
\end{figure}

There is a manifest equivalence between the CFT result in the above language, and the dual AdS calculation, as previously discussed 
in \cite{Hartman:2006dy,Giombi:2011ya}.  The crucial fact in proving this equivalence is 
the following identity for the difference of bulk-to-bulk propagators with standard and alternate boundary conditions, shown pictorially in Figure \ref{fig4}:
\begin{equation}
G_{d-\Delta}(z,\vec{x},w,\vec{y})-G_{\Delta}(z,\vec{x},w,\vec{y})
= 
\int d^d \vec{x}_0 d^d \vec{y}_0 K_{\Delta}(z,\vec{x};\vec{x}_0)K_{\Delta}(w,\vec{y};\vec{y}_0)
\frac{C_\sigma}{|\vec{y}_0-\vec{x}_0|^{2(d-\Delta)}}
\label{conv2}
\end{equation}
$C_\sigma$ was defined earlier as the normalization of the $\sigma$ propagator, eq. (\ref{sig-prop}). Here and elsewhere in this subsection we use the notation $x\equiv (z,\vec{x})$, $y\equiv (w,\vec{y})$ to denote points in AdS$_{d+1}$ in the 
usual Poincare coordinates, with $\vec{x}$,$\vec{y}$ etc. denoting coordinates of the flat boundary at $z=0$.  
In this identity $G_{\Delta}(z,\vec{x};w,\vec{y})$ is the bulk-to-bulk propagator with its canonical normalization, i.e. satisfying 
\begin{equation}
\left(-\nabla^2+m^2\right)G(x,y) = \delta^{(d+1)}(x,y)\,,\qquad m^2=\Delta(\Delta-d)\,,
\label{Green}
\end{equation}
and the bulk-to-boundary propagator is normalized as
\begin{equation}
K_{\Delta}(z,\vec{x};\vec{x}_0) = {\cal C}_{\Delta} \left(\frac{z}{z^2+(\vec{x}-\vec{x}_0)^2}\right)^{\Delta}\,,\qquad 
\C_{\Delta}=\frac{\Gamma(\Delta)}{2\pi^{d/2}\Gamma(\Delta+1-d/2)}
\end{equation}
Let us recall that with this choice of normalization of the bulk-to-boundary propagator, the two-point function of the dual operator is normalized as $\langle O (x_1)O (x_2)\rangle = {\cal C}_{\Delta}/x_{12}^{2\Delta}$. 

It is straightforward to prove \eqr{conv2} by assembling some known ingredients. First, one starts with the following single-integral identity \cite{Hartman:2006dy, Leonhardt:2003qu}:
\begin{equation}
G_{d-\Delta}(z,\vec{x};w,\vec{y})-G_{\Delta}(z,\vec{x};w,\vec{y})
=(2\Delta-d)\int d^d \vec{x}_0 K_{\Delta}(z,\vec{x};\vec{x}_0)K_{d-\Delta}(w,\vec{y};\vec{x}_0)\,.
\label{conv}
\end{equation}
Next, we use the following relation between the bulk-to-boundary propagators of conjugate dimension, which can be obtained by straightforward integration
\begin{equation}
K_{d-\Delta}(z,\vec{x};\vec{x}_0) =-\frac{\Gamma(d-\Delta)}{\pi^{\frac{d}{2}}\Gamma(d/2-\Delta)}
\int d^d\vec{y}_0 K_{\Delta}(z,\vec{x};\vec{y}_0)\frac{1}{|\vec{y}_0-\vec{x}_0|^{2(d-\Delta)}}
\label{K-conv}
\end{equation}
From this formula we see that changing the boundary conditions on bulk-to-boundary lines essentially amounts, from CFT point of view, to attaching the propagator of the 
auxiliary field $\sigma$. We can use (\ref{K-conv}) to rewrite (\ref{conv}) in the form \eqr{conv2}; we have used the fact that in this AdS calculation we can identify $C_{OO}=\C_{\Delta}$. We make some further remarks on the relation of these identities to AdS harmonic functions in Appendix \ref{appa}.

Given \eqr{conv2}, the relation to the CFT calculation of the ``two-triangle" diagram is transparent. The difference between the two exchange Witten diagrams in the channel $12\rightarrow 34$ for different boundary conditions on the intermediate field is, using (\ref{conv2}),
\begin{equation}
\begin{aligned}
\int d^d \vec{x}_0 d^d \vec{y}_0 &\left[\l_{\Phi\Phi O}\int dz d^d\vec{x} K_{\Delta_{\Phi}}(z,\vec{x};\vec{x_1})K_{\Delta_{\Phi}}(z,\vec{x};\vec{x_2})K_{\Delta}(z,\vec{x};\vec{x_0})\right]
\times \frac{C_\sigma}{|\vec{y}_0-\vec{x}_0|^{2(d-\Delta)}}\\
\times &\left[\l_{\Phi\Phi O}\int dw d^d\vec{y} K_{\Delta_{\Phi}}(w,\vec{y};\vec{x_3})K_{\Delta_{\Phi}}(w,\vec{y};\vec{x_4})K_{\Delta}(w,\vec{y};\vec{y_0})\right]\,.
\label{AdS4pt}
\end{aligned}
\end{equation}
where $\l_{\P\P O}$ is the AdS cubic vertex. The AdS integrals in brackets, each involving three bulk-to-boundary propagators, just give the three-point functions in the UV:
\begin{equation}
g_{\Phi\Phi O} \int dz d^d\vec{x} K_{\Delta_{\Phi}}(z,\vec{x};\vec{x_1})K_{\Delta_{\Phi}}(z,\vec{x};\vec{x_2})K_{\Delta}(z,\vec{x};\vec{x_0}) =
 \frac{C_{\Phi\Phi O}}{|\vec{x}_{12}|^{2\Delta_{\Phi}-\Delta} |\vec{x}_{20}|^{\Delta}|\vec{x}_{01}|^{\Delta}}
\end{equation}
in a normalization where $C_{\Phi\Phi}=\C_{\Delta_{\Phi}}$, and $C_{OO}=\C_{\Delta}$. Thus, we see that (\ref{AdS4pt}) precisely reproduces the ``two-triangle" diagram (\ref{2-triangle}) in the corresponding channel. Let us emphasize that our calculation does not rely on strong coupling, rather, only on large $N$ factorization and conformal symmetry.

\subsection{Relation to $SO(d+1,1)$ harmonic analysis}\label{harm}
The result \eqr{singchannel} for the difference in four-point functions in a single channel should equal a difference of conformal blocks for $O$ exchange in that same channel. This is clear from the AdS perspective explained in the previous subsection, combined with the fact that, in the {\it direct} channel conformal block decomposition of an AdS exchange diagram, the double-trace OPE data depends only on the squared mass, not the quantization, of the exchanged operator (e.g. \cite{Hijano:2015zsa}). Specifically, accounting for the difference in UV and IR OPE coefficients, we want to check that
\e{Ocheck}{{ a^{IR}_{\Phi\Phi O}\o a^{UV}_{\P\P O}}G_{d-\D,0}(u,v) - G_{\D,0}(u,v) = -\frac{\Gamma (\Delta )}{\Gamma ^4\left(\frac{\Delta }{2}\right) \Gamma \left(\frac{d}{2}-\Delta \right)}u^{\frac{d-\Delta}{2}}\bar{D}_{\frac{d-\Delta}{2},\frac{d-\Delta}{2},\frac{\Delta}{2},\frac{\Delta}{2}}(u,v)}
where $G_{\D,J}(u,v)$ is the conformal block for exchange of a dimension-$\D$, spin-$J$ operator. Note first that neither side depends on $\D_\P$. For the conformal blocks, we use the series representation \cite{Dolan:2000ut},
\e{}{g_{\D,0}(u,v) = \sum_{m,n=0}^\i\frac{ \left(\frac{\Delta }{2}\right)^2_m \left(\frac{\Delta }{2}\right)^2_{m+n}}{m! n! \left(-\frac{d}{2}+\Delta +1\right)_m (\Delta )_{2 m+n}}u^m(1-v)^n}
where $G_{\D,0}(u,v) = u^{\D/2}g_{\D,0}(u,v)$. For the $\Db$ function, we also use the series representation \cite{Dolan:2000ut}, which we can write as
\e{}{u^{\frac{d-\Delta}{2}}\bar{D}_{\frac{d-\Delta}{2},\frac{d-\Delta}{2},\frac{\Delta}{2},\frac{\Delta}{2}}(u,v) =\sum_{m,n=0}^\i{u^{\D\o2} \bar f_{mn}(\D) - u^{d-\D\o2}\bar f_{mn}(d-\D)\o m!n!}u^m(1-v)^n}
with
\e{}{\bar f_{mn}(\D) \equiv  \pi  \csc \left(\frac{\pi}{2} (d-2 \Delta )\right)\frac{\Gamma^2 \left(m+\frac{\Delta }{2}\right) \Gamma^2 \left(m+n+\frac{\Delta }{2}\right)}{\Gamma \left(-\frac{d}{2}+m+\Delta +1\right) \Gamma (2 m+n+\Delta )}}
Accounting for the ratio $a^{IR}_{\P\P O}/a^{UV}_{\P\P O}$ given in \eqr{aUV-aIR}, one finds agreement term-wise.

Now, conformally-covariant four-point functions may be expanded in a complete basis of single-valued functions living in the principal series representations of the conformal group, with $\D= {d\o2}+i\nu$, where $\nu\geq 0$ and is real (see e.g. \cite{Hogervorst:2017sfd, Hogervorst:2017kbj, Gadde:2017sjg, Caron-Huot:2017vep, Murugan:2017eto, Simmons-Duffin:2017nub} and references therein). In the notation of \cite{Simmons-Duffin:2017nub}, let us call these single-valued functions $\Psi_{\D,J}(x_i)$. These functions are not the ordinary conformal partial waves for unitary representations $\D\geq d-2+J$ with $\D\in\R$, which are not single-valued. The $\Psi_{\D,J}(x_i)$ may be written in terms of the conformal blocks as
\e{}{\Psi_{\D,J}(x_i) = {K_{d-\D,J}\o x_{12}^{2\DP}x_{34}^{2\DP}}\left(G_{\D,J}(u,v) + {K_{\D,J}\o K_{d-\D,J}}G_{d-\D,J}(u,v)\right)}
where
\e{}{K_{\D,J} \equiv {\pi^{d\o2}\G\left(\D-{d\o2}\right)\G\left(\D+J-1\right)\G^2\left({d-\D+J\o2}\right)\o \G\left(\D-1\right)\G\left(d-\D+J\right)\G^2\left({\D+J\o2}\right)}}
Now, we note that
\e{}{{K_{\D,0}\o K_{d-\D,0}} = -{a_{\Phi\Phi O}^{IR}\o a_{\Phi\Phi O}^{UV}}}
where $a_{\Phi\Phi O}^{IR}/a_{\Phi\Phi O}^{UV}$ was computed in \eqr{aUV-aIR}. Therefore, from \eqr{Ocheck}, we may write $\Psi_{\D,0}(x_i)$ simply in terms of a single $\Db$-function:
\e{}{\boxed{\Psi_{\D,0}(x_i) = {1\o x_{12}^{2\DP}x_{34}^{2\DP}}{\pi^{d\o2}\o \Gamma^2 \left(\frac{\Delta }{2}\right)\Gamma^2 \left(\frac{d-\Delta }{2}\right)}u^{\frac{d-\Delta}{2}}\bar{D}_{\frac{d-\Delta}{2},\frac{d-\Delta}{2},\frac{\Delta}{2},\frac{\Delta}{2}}(u,v)}}
This is a concise expression, valid for any $d$. We are giving both a mathematical expression for $\Psi_{\D,0}(x_i) $, as well as a physical interpretation in terms of double-trace RG flows. In AdS language, $\Psi_{\D,J}(x_i)$ is, up to a prefactor, equal to the difference of exchange diagrams for which the exchanged operator takes $\D_+$ and $\D_-$ quantizations. That this is true (for any $J$) is obvious from the identities of the previous subsection and Appendix \ref{appa}: CFT harmonic functions are dual to AdS harmonic functions. 

\ssec{Generalization to pairwise identical operators}
\label{pair-4pt}

The CFT calculation of the two-triangle diagram can be straightforwardly generalized to the case of four different external operators. With an eye toward applications, we explicitly consider here the case of pairwise identical scalar operators: given two scalar primary operators $\Phi$ and $\Psi$, we may compute 
the difference in the connected four-point function $\la \P\Psi\P\Psi\ra$ between UV and IR fixed points,
\e{}{\d\la\P(x_1)\Psi(x_2)\P(x_3)\Psi(x_4)\ra = {C_{\P\P} C_{\Psi\Psi}\o x_{12}^{\D_\P+\D_\Psi}x_{34}^{\D_\P+\D_\Psi}} \left(\frac{x_{24}}{x_{13}}\right)^{\DP-\D_\Psi}\d\F(u,v)}
Defining the norm-invariant ratio
\e{}{a_{\P\Psi O}^{UV} \equiv {C_{\Phi\Psi O}^2\o C_{\P\P}C_{\Psi\Psi}C_{OO}}}
the result in the $\P\Psi \rar \P\Psi$ channel is
\es{pair}{\d\F(u,v) &= -{a_{\P\Psi O}^{UV}\G(\D)\o \G^2\left({\D+\DP-\D_{\Psi}\o 2}\right) \G^2\left({\D-\DP+\D_{\Psi}\o 2}\right) \G\left({d\o 2}-\D\right)}u^{d-\D\o 2}\Db_{\D_1\D_2\D_3\D_4}(u,v) }
where
\es{}{\D_1 &= {d-\D+\D_{\Psi}-\D_{\Phi}\o 2}\\
\D_2 &= {d-\D-\D_{\Psi}+\D_{\Phi}\o 2}\\
\D_3 &= {\D+\D_{\Psi}-\D_{\Phi}\o 2}\\
\D_4 &= {\D-\D_{\Psi}+\D_{\Phi}\o 2}}
In the event that $\la \P\P O\ra=0$ or $\la \Psi\Psi O\ra=0$ -- for instance, if $O$ is neutral under a global symmetry under which $\Phi$ and $\Psi$ are charged -- the full, crossing-symmetric result is a sum of two terms,\footnote{See Appendix \ref{nonzerochannel} for the most general result in which these OPE coefficients are nonzero; there is simply one more term, of the same functional form as \eqr{pair}.} one in each available channel:
\es{pair2}{\d\F(u,v) &= -{a_{\P\Psi O}^{UV}\G(\D)\o \G^2\left({\D+\DP-\D_{\Psi}\o 2}\right) \G^2\left({\D-\DP+\D_{\Psi}\o 2}\right) \G\left({d\o 2}-\D\right)}\\&\times \Big(u^{d-\D\o 2}\Db_{\D_1\D_2\D_3\D_4}(u,v)  +\left({u\o v}\right)^{\D_\P+\D_{\Psi}\o 2}v^{d-\D\o 2}\Db_{\D_3\D_2\D_1\D_4}(u,v) \Big)}
Note that this is symmetric under $\DP \leftrightarrow \D_\Psi$, as it must be. This follows from the $\bar D$-function relations in Appendix \ref{DHG}.

As in Section \ref{sec2}, one can also derive the ratio between UV and IR three-point coefficients; the result is
\e{airpair}{a^{IR}_{\P \Psi O}= \frac{\G^2\left(\frac{d-\D -\DP + \D_\Psi}{2}\right) \G^2\left(\frac{d-\D +\DP - \D_\Psi}{2}\right) \G\left( \D - \frac{d}{2}+1 \right) \G(\D)}{\G^2\left(\frac{\D +\DP - \D_\Psi}{2}\right) \G^2\left(\frac{\D -\DP + \D_\Psi}{2}\right) \G\left(\frac{d}{2}-\D+1\right) \G\left(d-\D\right)} a^{UV}_{\P \Psi O}}

\sssec{Extremal case: $\D+\DP = \D_\Psi-2p$ or $\D+\D_\Psi = \DP-2p$}
\label{cancellation}
In these cases, we have zeroes from the gamma functions in \eqr{pair2} for all $p\in \Z_{\geq 0}$. Consequently, the result simplifies even further, which we now show for $p=0$. (See Appendix \ref{nonzerop} for the $p=1$ result and comments on general $p$.)

Suppose that $\D+\D_\Psi = \DP$. (The result for $\D+\DP=\D_\Psi$ is the same with $\D_\Psi \leftrightarrow\DP$.) First, note that the three-point coupling \eqr{airpair} vanishes in the IR, due to the $\G^{-2}(0)$ factor:
\e{}{a^{IR}_{\P\Psi O}=0}
This happens because for $\D+\D_\Psi = \DP$ -- indeed, for $\D+\D_\Psi = \DP-2p$ for $p\in\Z_{\geq 0}$ -- the UV three-point function is ``extremal'' \cite{DHoker:1999jke}. As explained in \cite{Giombi:2017mxl}, changing boundary conditions on an operator in an extremal correlator implies its vanishing at the new fixed point. 
Likewise, the prefactor in \eqr{pair2} vanishes. In order to extract the result for $\d \F$, we must introduce a regulator. Take $\D+\D_\Psi=\DP + 2\eps$. Then
\e{}{\D_1 = {d\o 2}-\D + \eps ~, ~~ \D_2 = {d\o 2} - \eps ~, ~~ \D_3 = \eps~, ~~ \D_4 = \D-\eps}
To determine the finite piece of \eqr{pair2}, we must extract the $\O(\eps^{-2})$ term in the $\Db$-functions, to cancel the $\eps^2$ prefactor. Using their double-sum representation (see Appendix \ref{DHG}), one simply finds\footnote{In particular, everything other than the $\G^2(\eps)$ is regular when $\eps \rightarrow 0$, including the $G$-functions \eqref{G}.}
\e{332}{\Db_{{d\o 2}-\D+\eps,{d\o 2}-\eps,\eps,\D-\eps}(u,v) \approx  \frac{u^{\Delta -{d\o 2}}\G({d\o 2}-\D)\G(\D) }{\epsilon ^2}+O(\eps^{-1}) }
Combining this with \eqr{pair2}, we get
\e{degfinal}{\d\F(u,v) \approx -{a^{UV}_{\P\Psi O}}\, u^{\Delta\o 2}\Big(1  +\left({u\o v}\right)^{\D_{\Psi}}\Big)}
This is the final, extremely simple, result. Remarkably, \eqr{degfinal} is $d$-independent, and is manifestly negative for all real $u,v$. We will soon see other sign-definite properties of the double-trace RG flow.

\section{Microscopics: Extracting OPE data}\label{s4}
Having derived the change in a connected four-point function along the double-trace flow to leading order in $1/N$, we may extract the change in OPE data by branching it into conformal blocks. Under this deformation, the single-trace spectrum is identical between UV and IR to leading order in $1/N$, except for the dimension of $O$. However, the {\it double-trace} contributions to the leading-order connected correlator also are modified. That this is true can be seen by considering the requirement of crossing symmetry: if only the $O$ exchange is modified, this will spoil crossing symmetry unless we compensate with changes in the other operator exchanges. Because this is a connected correlator at leading order in $1/N$, the only other exchanges are double-trace operators.

We focus for now on the result (\ref{4pt-final}) for the case of identical external operators. In general, the four-point function has the conformal block expansion
\begin{equation}
\begin{aligned}
&\langle\Phi(x_1)\Phi(x_2)\Phi(x_3)\Phi(x_4) \rangle = 
\frac{C_{\Phi\Phi}^2}{x_{12}^{2\Delta_{\Phi}} x_{34}^{2\Delta_{\Phi}}}{\cal F}(u,v)\\
& {\cal F}(u,v) = \sum_{\tau,s} a_{\tau,s} u^{\frac{\tau}{2}} g_{\tau,s}(u,v)
\label{genOPE}
\end{aligned}
\end{equation}
$a_{\t,s}$ are the normalized squared OPE coefficients, and $u^{\t\o 2}g_{\t,s}(u,v)$ is the conformal block for exchange of a conformal primary of twist $\tau$ and spin $s$, where twist is defined as $\t=\D-s$. The double-trace primaries \eqr{dtprim} have twists and OPE coefficients that admit a $1/N$ expansion,\footnote{Recall that we define $N$ as $c_T \sim N$.}
\es{}{\anl -\anl^{(0)}&\approx {1\o N}\anl^{\1}+\ldots\\
\t_{n,\ell} - \t_{n}^{(0)} &\approx {1\o N} \gnl+\ldots}
where $\t^{(0)}_{n} = 2\D_\P+2n$, and the mean field theory $(N=\i)$ OPE coefficients $a^{(0)}_{n,\ell}$ are known in general $d$ \cite{Fitzpatrick:2011dm}. This, in turn, induces a $1/N$ expansion of $\F(u,v)$ at each fixed point. Taking the difference of IR and UV connected correlators to leading order in $1/N$,
\e{dF}{\d \F(u,v) = \d \F^{O}(u,v) + \d \F^{[\P\P]}(u,v)}
where
\e{dFO}{\d \F^{O}(u,v)  \equiv a^{IR}_{\Phi\Phi O}u^{d-\D\o 2}g_{d-\D,0}(u,v) - a^{UV}_{\Phi\Phi O}u^{\D\o 2}g_{\D,0}(u,v)}
and
\e{dFPP}{\d\F^{[\Phi\Phi]}(u,v) \equiv u^{\D_{\Phi}}\sum_{n=0}^\i \sum_{\ell=0,2,\ldots}^\i \left({1\o2}\anl^{\0}\d\gnl\p_{n} + \d\anl^{(1)}\right)u^n g_{\t_n^{(0)},\ell}(u,v)}
Writing \eqr{4pt-final} in the form \eqr{dF}--\eqr{dFPP} and expanding in powers of $u$ and $1-v$, we can extract all OPE data. 

In Section \ref{harm}, we showed that the $O$ exchange piece $\d \F^{O}(u,v)$ is indeed accounted for by the first term in \eqr{4pt-final}, i.e. the direct-channel term. In the rest of this section, we focus on the double-trace terms \eqr{dFPP}, which come from crossed-channel contributions and contain interesting data.\footnote{A complete way to perform these calculations is to use Caron-Huot's inversion formula \cite{Caron-Huot:2017vep}. This requires taking a double-discontinuity of our $\Db$-function. We will instead use more basic tools.}

\ssec{Double-trace anomalous dimensions}
First we extract $\d \gnl$, focusing in particular on the leading-twist tower $n=0$.   (Higher $n$ may be computed in a systematic expansion in small $u$.) The final result is in \eqr{dg0ell}. In this subsection and elsewhere, whenever we focus on the $n=0$ tower for some double-trace observable $X$, we use the notations
\e{}{\d X_\ell \equiv \d X_{0,\ell}}

We proceed by isolating the $\log u$ piece of the full double-trace contribution
\e{dgceq}{\d\F^{[\Phi\Phi]}(u,v) = u^{\D_{\Phi}}\sum_{n=0}^\i u^n\sum_{\ell=0,2,\ldots}^\i \left({1\o2}\anl^{\0}\d\gnl\log u + \d\anl^{(1)}+{1\o2}\anl^{\0}\d\gnl\p_n \right) g_{\t_n,\ell}(u,v)}
In what follows, we will concentrate on $\d \g_{\ell}$. So to extract the anomalous dimensions, we will need the expansion of the $\log u$ terms of $\bar{D}$-functions in (\ref{4pt-final}) in the OPE limit. These terms are given by\footnote{Note that the $\bar{D}_{\frac{d-\Delta}{2},\frac{d-\Delta}{2},\frac{\Delta}{2},\frac{\Delta}{2}}(u,v)$ term in \eqr{4pt-final} does not contain $\log u$ terms (for 
generic $\Delta$ and $\Delta_{\Phi}$), consistent with the fact that it contributes only to $\d\cF^O$, as shown in Section \ref{harm}, and not to $\d\cF^{[\P\P]}$.} 
\begin{equation}
\begin{aligned}
&\bar{D}_{\frac{d-\Delta}{2},\frac{\Delta}{2},\frac{d-\Delta}{2},\frac{\Delta}{2}}(u,v)\big|_{\log u} = 
- \frac{\Gamma^2 \left(\frac{\Delta }{2}\right)\Gamma^2 \left(\frac{d-\Delta }{2}\right)}{\Gamma \left(\frac{d}{2}\right)}
G\left(\frac{\Delta }{2},\frac{d-\Delta }{2},1,\frac{d}{2};u,1-v\right)\\
&\bar{D}_{\frac{\Delta}{2},\frac{d-\Delta}{2},\frac{d-\Delta}{2},\frac{\Delta}{2}}(u,v)\big|_{\log u} = 
- \frac{\Gamma^2 \left(\frac{\Delta }{2}\right)\Gamma^2 \left(\frac{d-\Delta }{2}\right)}{\Gamma \left(\frac{d}{2}\right)}
G\left(\frac{d-\Delta }{2},\frac{d-\Delta }{2},1,\frac{d}{2};u,1-v\right)
\label{Dbar-OPE}
\end{aligned}
\end{equation}
where the $G$-function, introduced by Dolan and Osborn \cite{Dolan:2000uw}, admits the following double-series expansion:
\e{}{G(\alpha,\beta,\gamma,\delta;u,1-v)=\sum_{n,m=0}^{\infty} 
\frac{(\delta-\alpha)_m (\delta-\beta)_m}{m! (\gamma)_m} 
\frac{(\alpha)_{m+n} (\beta)_{m+n}}{n! (\delta)_{2m+n}} 
u^m (1-v)^n}
Note that the $G$-function obeys a small-$u$ expansion
\e{g2f1}{G(\a,\b,1,\d;u,1-v) = {}_2F_1(\a,\b,\d;1-v) + \O(u)}
\sssec{$\Phi^2$}
Before giving a general result for $\d\g_{\ell}$, let us start by extracting $\d\g_0$, the change in the anomalous dimension of the leading-twist scalar operator $:\!\Phi^2\!:\,\,=[\Phi\Phi]_{0,0}$. In this case, 
we have $a_{\tau,\ell}=a_{2\Delta_{\Phi},0}=2+\O(1/N)$, and to leading order at small $u$ we just have 
$$
u^{\tau/2} g_{\tau,\ell}(u,v)=u^{\Delta_{\Phi}}(1+\frac{\gamma_{0}}{2}\log u+\ldots)(1+\ldots)\,.
$$ 
Then, using (\ref{Dbar-OPE}) into (\ref{4pt-final}) and matching to (\ref{genOPE}), we find the result
\begin{equation}\label{g00}
\boxed{\d\g_{0} = a^{\rm UV}_{\Phi\Phi O} \frac{2\Gamma (\Delta ) \Gamma^2 \left(\frac{d-\Delta }{2}\right)}{\Gamma \left(\frac{d}{2}\right) \Gamma^2\left(\frac{\Delta }{2}\right) \Gamma \left(\frac{d}{2}-\Delta \right)}\,.}
\end{equation}
We point out two features of this result. First, it is manifestly positive for all ${d-2\o 2}< \D < {d\o 2}$. Second, it is highly non-trivial that this does not depend on $\DP$, because $\g_{0}^{IR}$ and $\g_{0}^{UV}$ both do. This $\DP$-independence will not persist at higher $\ell$. Note that, using (\ref{aUV-aIR}), we can also express \eqr{g00} as
\begin{equation}
\d\g_{0}= 
a^{\rm UV}_{\Phi\Phi O} \frac{\Gamma (\Delta ) \Gamma^{2} \left(\frac{d-\Delta }{2}\right)}{ \Gamma \left(\frac{d}{2}\right) \Gamma^{2} \left(\frac{\Delta }{2}\right)\Gamma \left(\frac{d}{2}-\Delta \right)}
-a^{\rm IR}_{\Phi\Phi O}\frac{\Gamma (d-\Delta )\Gamma^{2} \left(\frac{\Delta }{2}\right)}{ \Gamma \left(\frac{d}{2}\right) \Gamma ^{2}\left(\frac{d-\Delta }{2}\right) \Gamma \left(\Delta -\frac{d}{2}\right)}
\end{equation}
which is manifestly odd under $\Delta \rightarrow d-\Delta$, as required.
\sssec{Leading-twist}\label{s422}
We now derive $\d\g_{\ell}$ in closed form. First, we introduce a notation
\e{kfunc}{F_{\b}(z) \equiv {}_2F_1(\b,\b,2\b,z)~.}
In the lightcone regime $u\ll1$, the conformal blocks become the collinear $SL(2,\mathbb{R})$ blocks,\footnote{This defines a convention for the conformal block normalization.}
\es{}{g_{\t_n,\ell}(u\ll1,v) \approx g_{n,\ell}^{coll}(v) = x^{\ell}F_{\ell}(x)~, ~~\text{where}~~ x\equiv 1-v~.}
To leading order at small $u$, the $\log u$ term of \eqref{dgceq} becomes
\e{logeq}{\d\F^{[\Phi\Phi]}(u,v)\Big|_{u^{\D_{\Phi}}\log u} \approx {1\o 2}\sum_\ell a_{\ell}^{\0}\d\g_{\ell}x^{\ell}F_{\ell}(x)}
Now, $x^\b F_\b(x)$ are eigenfunctions of the operator $D = x^2(1-x)\p_x^2 - x^2 \p_x$, with eigenvalue $\b(\b-1)$. 
%
%\e{}{D k_\b(x) = \b(\b-1)k_\b(x)~, \quad \text{where}~~D = x^2(1-x)\p_x^2 - x^2 \p_x}
%
They obey the orthogonality condition
\e{proj}{{1\o 2\pi i}\oint_{x=0} {x^{\b-\b'-1}} F_\b(x) F_{1-\b'}(x) = \delta_{\b,\b'}}
where $\b-\b'\in\Z$, and the contour runs counterclockwise around the origin. This was used in a similar context in e.g. \cite{Heemskerk:2009pn, Alday:2017gde}. Applying this to \eqr{logeq},
\e{gameq}{\d\g_{\ell} = {1\o \pi i a_{\ell}^{\0}}\oint_{x=0}x^{-1-\ell}F_{1-\D_{\Phi}-\ell}(x)\, \left[\d\F^{[\Phi\Phi]}(u,1-x)\big|_{u^{\D_{\Phi}}\log u}\right] }
which is the desired result. % I checked against a handful of examples obtained by brute force expansion at various fixed $d,\D_{\Phi}$. 
Actually, we may go further and explicitly extract the residue in closed form: from \eqr{g2f1}, it is clear that we need only isolate a term of order $x^\ell$ in the product of various hypergeometric functions. We carry this out in Appendix \ref{2F1}. The final result can be written as a finite sum:
\begin{equation}
\frac{a^{(0)}_{\ell}\delta \gamma_{\ell}}{a^{(0)}_{0} \delta \gamma_{0} }= \sum_{m=0}^{\ell}(-1)^m \frac{\left(\frac{\Delta }{2}\right)_m \left(\frac{d-\Delta }{2}\right)_m 
\left(\Delta _{\Phi }+m\right)^2_{\ell-m}}{m! (\ell-m)! \left(\frac{d}{2}\right)_m 
\left(\ell+2\Delta _{\Phi }+m-1\right)_{\ell-m}}
\label{gamell}
\end{equation}
where $a_{\ell}^{(0)}$ are the mean field theory OPE coefficients \cite{Heemskerk:2009pn, Fitzpatrick:2011dm}
\e{mftope}{a^{(0)}_{\ell} = \frac{(1+(-1)^{\ell}) (\DP)_\ell^2}{\ell! (2\DP +\ell -1)_{\ell}}}
This can be neatly written in terms of a terminating ${}_4F_3$  hypergeometric function, with no explicit appearance of $a^{\0}_{\ell}$:
\begin{equation}\label{dg0ell}
\boxed{\delta \gamma_{\ell} = 
 {}_4F_3\Farg{-\ell,\,\frac{d-\Delta}{2},\,\frac{\Delta }{2},\,2 \Delta _{\Phi }+\ell-1}{\frac{d}{2},\,\Delta _{\Phi },\,\Delta _{\Phi }}{1}\,
\delta \gamma_{0}}
\end{equation}
with $\d\g_{0}$ given in \eqr{g00}.

As a consistency check on this result, evaluating (\ref{dg0ell}) in the large $\ell$ limit, one finds
\begin{equation}\label{lcboot}
\delta\gamma_{\ell} \approx 
\frac{2 a^{\rm UV}_{\Phi\Phi O}  \Gamma (\Delta ) \Gamma^{2} \left(\Delta _{\Phi }\right)}{\Gamma ^{2}\left(\frac{\Delta }{2}\right) \Gamma^{2} \left(\Delta _{\Phi }-\frac{\Delta }{2}\right)} \frac{1}{\ell^{\Delta}}
-\frac{2  a^{\rm IR}_{\Phi\Phi O} \Gamma (d-\Delta ) \Gamma^2 \left(\Delta _{\Phi }\right)}{\Gamma^{2} \left(\frac{d-\Delta }{2}\right)
\Gamma^{2}\left(\Delta _{\Phi }-\frac{d-\Delta}{2}\right)} \frac{1}{\ell^{d-\Delta}}+\ldots 
\end{equation}
This correctly reproduces the leading-order results of the lightcone bootstrap  \cite{Komargodski:2012ek,Fitzpatrick:2012yx}: it is a difference of the leading large-spin asymptotics of $\d\g_{\ell}$ in the IR and UV.

\sssec{Comments}

\sssec*{RG monotonicity}
As we noted earlier, it is interesting that $\d\g_{0}$ in \eqr{g00} is always positive under double-trace RG flows: that is, $\g_{0}$ is {\it greater} in the IR than the UV,
\e{}{\g_{0}^{IR} \geq \g_{0}^{UV}~,}
for any spacetime dimension $d$. 

What about $\d\g_{\ell}$ for higher spins? At large spin, $\ell^\D\ll \ell^{d-\D}$, so the UV term of \eqr{lcboot} dominates (assuming $a^{UV}_{\P\P O}\neq 0$), implying $\d\g_{\ell \gg 1}>0$ due to the positivity of $a^{UV}_{\P\P O}$. Thus, any negativity must be confined to finite $\ell$. By a combination of analytical arguments and numerically sampling many values of parameters, we find the following condition:
\e{d23mon}{\g^{IR}_{\ell}\geq \g^{UV}_{\ell} ~~\text{when}~~\DP\geq {d\o 4}\,.}
This implies that 
\e{d4mon}{\g^{IR}_{\ell}\geq \g^{UV}_{\ell} ~~\text{for}~~ d\geq 4\,,}
for all unitary values of parameters.

\eqr{d4mon} can be proven in $d=4$ as follows. Due to the $\D \rar d-\D$ symmetry of the ${}_4F_3$, its extremum as a function of $\D$ sits at $\D={d\o 2}$; one can check that it is a minimum. Taking $d=4$, we now utilize the identity \cite{Prudnikov} (see p.470, eq. 46)
\e{}{{}_4F_3\Farg{-\ell, 1,1, 2\DP+\ell-1}{2,\DP,\DP}{1} = {2(\DP-1)^2\o (\ell+1)(2\DP+\ell-2)}\Big[\psi(\DP+\ell)-\psi(\DP-1)\Big]}
where $\psi$ is the digamma function. The prefactor is manifestly positive for all $\DP>1$ and $\ell>0$. The difference of digamma functions is also positive, because $\psi'(x)>0$ for $x>0$. Therefore, $\d\g_{\ell}$ is indeed positive under double-trace RG flow. For $d\neq 4$, one can show that at $\D=d/2$ and $\DP=d/4$, $\d\g_{\ell}$ has a zero for all $\ell\in \mathbb{Z}$, because
\es{}{{\d\g_{\ell}\o\d\g_{0}}\Bigg|_{\substack{\D=d/2,\\ \DP=d/4}}%= {}_4F_3\Farg{-\ell,\,\frac{d}{4},\,\frac{d}{4},\,{d\o 2}+\ell-1}{\frac{d}{2},\,{d\o4},\,{d\o4}}{1} 
= {\G\left({d\o2}\right)\o \G(1-\ell)\G\left({d\o2}+\ell\right)}}%\quad \quad (\D=d/2, ~ \DP=d/4)}
Then by plotting many values, one sees that \eqr{d23mon} holds. 
 \begin{figure}
    \centering
       \includegraphics[width = .65\textwidth]{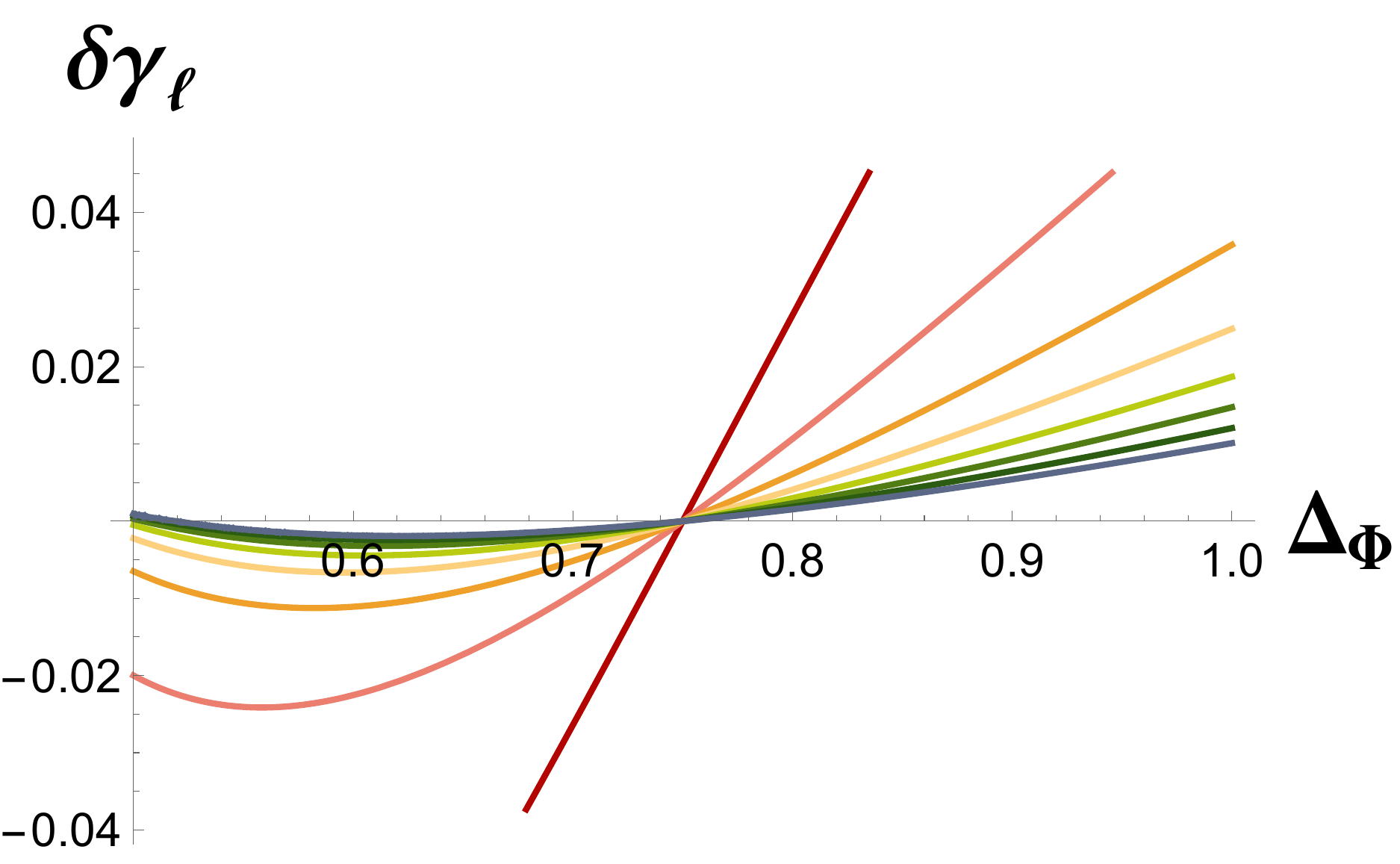}
         \caption{In $d=3$, a plot of $\d\g_{\ell}$ evaluated at $\D=3/2$, its minimum, as a function of $\DP$. We plot $\ell=2,4,\ldots,16$, where red is $\ell=2$ and the spin increases as we move through the rainbow. For $\DP>3/4$, the function is positive for all $\ell$.}\label{fig5}
\end{figure}
 In Figure \ref{fig5}, we exhibit this behavior in $d=3$. This conclusion would follow if, as suggested by the sampling, $\DP=d/4$ is the only zero of $\d\g_{\ell}$ at $\D=d/2$, viewed as a function of $\DP$, for unitary values of $\DP$.

\sssec*{Flows from UV free CFTs}

One may be puzzled about cases in which the UV is a free theory, so that $\delta\gamma_{\ell}=\gamma^{IR}_{\ell}$. In such cases -- again assuming $a^{UV}_{\P\P O}\neq 0$ -- one has $\g^{IR}_{\ell\gg1}>0$, because the UV term of \eqr{lcboot} dominates . This conflicts with naive lightcone bootstrap intuition at large spin. The resolution to this is that the UV free theory contains an infinite tower of conserved higher spin currents which becomes nearly conserved in the IR, and negativity of $\gamma_{\ell}$ does not apply, even at large spin \cite{Alday:2015ota}: a resummation is required. What our result shows is that, in fact, {\it every} CFT with slightly broken higher-spin symmetry that is obtained by double-trace flow from a UV-free CFT has $\gamma^{IR}_{\ell\gg1}>0$.

\sssec*{Heavy operators}

Consider the result \eqr{dg0ell} for the change, under double-trace flow, of the anomalous dimensions of the double-trace operators $[\P\P]_{0,\ell}$. We now suppose the external operator $\P$ is a heavy operator, with $1\ll \DP \ll N$. Such an operator may be, for instance, a string-scale operator in a large $N$ gauge theory. 

First, suppose that $\ell$ remains finite as we dial $\DP\gg1$. In this case, $\d\g_{\ell} = \d\g_{0}$ to leading order in $1/\DP$. The reason for this is clear in AdS: the binding energy for a bound state of two heavy particles with $m L_{\rm AdS} \gg 1$ will be unaffected by the addition of a parametrically small angular momentum $J\ll m$. More interesting is the regime in which 
\e{}{\DP\gg1~, ~~ \ell \gg 1~, ~~ \eta\equiv {\ell\o \DP} ~\text{fixed}}
Representing the ${}_4F_3$ in series form and taking the limit of the summand, one obtains a finite result; performing the sum then yields an ordinary hypergeometric function,
\e{heavy}{\d\g_{\ell} \approx {}_2F_1\left({\D\o2},{d-\D\o2}, {d\o 2}, -\eta(\eta+2)\right)\d\g_{0}+\O(\DP^{-1})}
One readily confirms that for $\eta \rar\i$ -- that is, $1 \ll \DP\ll \ell$ -- we recover the $\DP\gg 1$ limit of the large spin expansion \eqr{lcboot}. It would be interesting to reproduce \eqr{heavy} from a bulk computation in which $\d\g_{\ell}$ is the difference in the contribution of $O$, for standard versus alternate quantizations, to the binding energy of the $[\P\P]_{0,\ell}$ state, where $\Phi$ is represented as a particle moving along a bulk worldline.

\ssec{Double-trace OPE coefficients}\label{dtope}

One can also derive the change in OPE coefficients, $\d a_{n,\ell}^{(1)}$, in \eqr{dFPP}. We eschew a comprehensive treatment here, only giving the lowest-lying contribution. Again starting from \eqr{4pt-final}, we use the form of the $\Db$-functions in Appendix \ref{DHG} -- in particular, equations \eqr{bard} and \eqr{a6} -- to obtain
\e{ope00}{\d a^{(1)}_{0} = 2\d\g_{0} \left(\psi \left(\frac{d-\Delta }{2}\right)-\psi \left(\frac{d}{2}\right)+\psi \left(\frac{\Delta }{2}\right)+\gamma_E \right)}
where $\g_E$ is the Euler constant. Note the interesting feature that, like $\d\g_{0}$, this is independent of $\DP$. There is simplification for various rational values of $\D,d$. One can continute iteratively for low $\ell$ as desired.

\ssec{Generalization to pairwise identical operators}\label{s44}
We now perform the same analysis for the pairwise identical correlator $\la \P\Psi\P\Psi\ra$, whose change under the RG flow was derived in \eqr{pair2}. 

Let us start by deriving the change in anomalous dimensions, $\d\g_{\ell}$, for the leading-twist double-trace operators $[\P\Psi]_{0,\ell}$. The strategy for the calculation is the same as for the correlator $\la \P\P\P\P\ra$: starting with the result for $\d\F(u,v)$ in \eqr{pair2}, we expand in collinear blocks and apply a projector. Some formulas become rather unwieldy, so we present the results here and describe the detailed calculation in the Appendix \ref{generalnonidentical}. We arrive at the following generalization of \eqref{gamell}:
\begin{equation}
\frac{a^{(0)}_{\ell}\delta \gamma_{\ell}}{a^{(0)}_{0} \delta \gamma_{0} }=  \sum_{m=0}^{\ell} (-1)^{m} \frac{\left(\frac{\Delta-\DP+\D_\Psi }{2}\right)_m \left(\frac{d-\Delta-\DP+\D_\Psi }{2}\right)_m 
\left(\Delta _{\Psi }+m\right)^2_{\ell-m}}{m! (\ell-m)! \left(\frac{d}{2}\right)_m 
\left(\ell+\Delta _{\Phi}+\D_\Psi+m-1\right)_{\ell-m}}
\label{gamelluneq}
\end{equation}
Using the explicit values of the infinite $N$ OPE coefficients $a^{(0)}_{\ell}$ from \cite{Fitzpatrick:2011dm},
\e{mftope2}{a^{(0)}_{\ell} = \frac{(-1)^{\ell} (\DP)_\ell (\D_\Psi)_\ell}{\ell! (\DP+\D_\Psi +\ell -1)_{\ell}}}
the sum above can be rewritten as\footnote{We thank Charlotte Sleight and Massimo Taronna \cite{Sleight:2018epi} for pointing out a typo in this formula in the first version of the paper.} 
\begin{equation}\label{pairdgam}
\boxed{\d \gamma_{\ell} =(-1)^{\ell}{(\D_\Psi)_\ell\o (\DP)_{\ell}}\,  {}_4 F_3\Farg{-\ell,\, \ell+\DP+\D_\Psi-1,\,\frac{d- \D -\DP +\D_\Psi}{2},\,\frac{\D-\DP+\D_\Psi}{2}}{\D_\Psi,\,\D_\Psi,\,\frac{d}{2}}{1} \d \gamma_{0}}
\end{equation}
where the spin-zero anomalous dimension is
\e{e9a}{\boxed{\delta \gamma_{0} =  a_{\P\Psi O}^{UV}{2 \G(\D) \G\left({d-\D+\DP-\D_{\Psi}\o 2}\right) \G\left({d-\D-\DP+\D_{\Psi}\o 2}\right) \o \G\left(d \o2 \right)\G\left({d\o 2}-\D\right)   \G\left({\D+\DP-\D_{\Psi}\o 2}\right) \G\left({\D-\DP+\D_{\Psi}\o 2}\right)}} }
Unlike the identical operator case, this is valid for odd integer $\ell$ as well. Note that if $\DP=\D_\Psi$, \eqr{pairdgam} reduces to the result \eqr{dg0ell} for identical operators. This may be confirmed by direct calculation.

This result must be symmetric under $\DP\leftrightarrow \D_\Psi$, because it captures coefficients in the expansion of the symmetric function \eqr{pair}, but this symmetry is not manifest in \eqr{pairdgam}. However, the result in \eqr{pairdgam} can be expressed in terms of a certain orthogonal polynomial, known in the literature as a Wilson polynomial \cite{wilson1980some}:\footnote{It is not clear whether the orthogonality property is physically interesting here. It is also interesting to note that these and the related ``Wilson functions'' have appeared recently in the physics context as fusion matrices in 2d CFT and as scattering phases near AdS black holes \cite{Hogervorst:2017sfd, Hogervorst:2017kbj, Mertens:2017mtv}. In the present context, the $\d\g_\ell$ should be thought of as 6j-symbols for the confomal group, which would follow from their derivation (not performed here) from the inversion formula \cite{Caron-Huot:2017vep}. We thank David Simmons-Duffin for discussion on this and related issues.}
\es{}{{\d\g_{\ell}\o\d\g_{0}} &=\frac{1}{ \left(\frac{d}{2}\right)_\ell \left(\DP\right)_\ell \left(\D_\Psi\right)_\ell}\\&\times p_\ell \left(\frac{2\D -d}{4}; \frac{d-\DP +\D_\Psi}{4},\frac{d+\DP -\D_\Psi}{4},\frac{\DP+\D_\Psi}{2}-{d\o4},\frac{\DP+\D_\Psi}{2}-{d\o4} \right)}
%\end{equation}
%
These are known to be symmetric in the last four arguments, which includes the transformation $\DP \leftrightarrow \D_\Psi$.

In parallel with section \eqr{dtope}, one may also derive, $\d a_{0}^{(1)}$, which is now the difference in (normalized) squared OPE coefficients $C_{\P\Psi[\P\Psi]_{0,0}}^2$; the result may be found in \eqr{a00nonidentical}. 

\sssec{Extremal case: $\DP=\D_\Psi+\D$}
Recall that for this extremal alignment of dimensions, we obtained the simple result in \eqr{degfinal}, where we assumed that the $\Phi\Phi \rar O \rar \Psi\Psi$ channel is absent. 

The first term in \eqr{degfinal} represents the UV exchange of $O$,
\es{}{\d\F^O(u,v) &=  - a^{UV}_{\Phi\Psi O}u^{\D\o 2}g_{\D,0}(u,v)\\& = -a^{UV}_{\P\Psi O}u^{\Delta\o 2}}
where we have used that $a^{IR}_{\Phi\Psi O}=0$ in the first line. The fact that the conformal block is simply equal to unity can be checked explicitly using the $d=4$ blocks,
\begin{equation}\label{4dcb}
g_{\t,\ell}(z,\zb) =  {1\o z-\zb}\left(z^{\ell+1} F^{(\D_{12},\D_{34})}_{\tau/2+\ell}(z) F^{(\D_{12},\D_{34})}_{\frac{\tau-2}{2}}(\bar z)-\bar z^{\ell+1} F^{(\D_{12},\D_{34})}_{\tau/2+\ell}(\bar z) F^{(\D_{12},\D_{34})}_{\frac{\tau-2}{2}}(z) \right)
\end{equation}
where
\e{}{F_\b^{(\D_{12},\D_{34})}(z) \equiv {}_2F_1\left(\b-{\D_{12}\o 2}, \b+{\D_{34}\o 2}, 2\b,z\right)}
When $\D_{12}=\D_{34}=\pm \D$, indeed one has $g_{\D,0}(z,\zb)=1$. 

The second term in \eqr{degfinal} represents the exchanges of $[\Psi\Phi]_{n,\ell}$:
\e{identicalfourspecial}{\d\F^{[\P\Psi]}(u,v) = -{a^{UV}_{\P\Psi O}}u^{\DP+\D_\Psi\o 2} v^{-\D_{\Psi}}}
The absence of a $\log u$ term implies that, consistent with \eqr{e9a},
\e{}{\d\g_{n,\ell}=0~,~~\forall~ (n,\ell)~.}
Moreover, due to the simple form of the result, we can derive explicit formulas for $\d a^{(1)}_{\ell}$. In terms of conformal blocks, \eqr{identicalfourspecial} comes from a sum, over all $n,\ell$, due to the change in OPE coefficients $\d a^{\1}_{n,\ell}$: in particular, they obey the sum rule
\e{440}{\sum_{n,\ell}^\i \d a^{(1)}_{n,\ell}u^n g_{\t_n^{(0)},\ell}(u,v) = -{a^{UV}_{\P\Psi O}}v^{-\D_\Psi}}
The right-hand side is independent of $u$. This implies that if we expand the left-hand side in powers of $u$, we obtain an infinite set of equations. First, at zeroth order,
\e{a0lpair}{\sum_{\ell=0,2,\ldots}^\i \d a^{\1}_{\ell}g^{coll}_{0,\ell}(v) =-{a^{UV}_{\P\Psi O}} v^{-\D_\Psi}}
where $g^{coll}_{0,\ell}(u,v)$ is the $n=0$ collinear block for pairwise identical operators in the $\Phi\Psi \rar \Phi\Psi$ channel, defined in \eqr{gcollpair}. Since \eqr{a0lpair} is a sum over collinear blocks, it can be solved for $\d a^\1_{\ell}$ by using the projector for the collinear blocks, just as for the $\d\g_{\ell}$ in Appendix \ref{generalnonidentical}. %
The result is
\e{}{{\d a^\1_{\ell} = -{a^{UV}_{\P\Psi O}}\frac{ (\DP+\ell)_\ell (\D_\Psi+\ell)_\ell}{\ell!(\DP+\D_\Psi+2\ell)_\ell}\, _3F_2\Farg{-\ell,\D_\Psi,-\DP-\D_\Psi-3 \ell+1}{-\DP-2 \ell+1,-\D_\Psi-2 \ell+1}{1}}}
As a check, this agrees with the \eqref{a00nonidentical} derived for general $\D_\P$, $\D_\Psi$, when specialized to the case $\DP=\D_\Psi+\D$. Note that at spin-zero,
\e{ope00deg}{\d a^\1_{0} = -{a^{UV}_{\P\Psi O}}}
which is much simpler than the spin-zero result for identical operators in \eqr{ope00}.  

Moreover, all terms in \eqr{440} carrying powers of $u$ must vanish. This is allowed by unitarity because there is no sign constraint on the $\d a^{(1)}_{n,\ell}$ (nor on the individual $a^{(1)}_{n,\ell}$, which are $1/N$--suppressed compared to $a^{(0)}_{n,\ell}$). Solving the resulting infinite set of equations would yield $\d a_{n,\ell}^{(1)}$. 

\ssec{Adding global symmetries}

If $\P$ and $O$ carry charges under some global symmetry group $\Gc$, this requires a slight modification of our formulas. Let us call the exchanged operator $O^I$, where $I$ indicates that the operator sits in some representation of $\Gc$. There are various possible double-trace deformations that involve some subset of components of $O^I$. The most symmetric choice is to activate the singlet,
\e{}{S_\l = S_{CFT} + \l\int d^d x \,O_IO^I}
which preserves $\Gc$. To analyze the IR CFT, we introduce a Hubbard-Stratonovich field $\sigma_I$, which couples as $\int d^d x\, \sigma_I O^I$ and carries the same charges as $O^I$. Correlation functions of these operators now carry extra dependence on the representations involved, but the calculations are otherwise essentially identical. 

In particular, let us take $\P$ and $O$ to sit in representations $\Rc_\P$ and $\Rc_O$ of $\Gc$, respectively, where $\Rc_\P \otimes \Rc_\P \supset \Rc_O$. The three-point functions of Section \ref{sec2} are unchanged, up to an overall tensor that encodes this product of representations. Similarly, we can return to the calculation of the change in the four-point function, \eqr{4pt-final}. The result in a single channel is the same as \eqr{singchannel}, up to multiplication by a tensor ${\cal T}^{\Rc_O}_{1234}$ that accompanies the exchange of $O$, where the subscript labels the external points. Adding the three channels together yields the total result,
\e{global}{\d\F(u,v)={\cal T}_{1234}^{\Rc_O}\times u^{\frac{d-\Delta}{2}}\bar{D}_{\frac{d-\Delta}{2},\frac{d-\Delta}{2},\frac{\Delta}{2},\frac{\Delta}{2}}(u,v)+\text{perms}}
where we must permute the indices of ${\cal T}^{\Rc_O}_{1234}$ as well as the positions of the operators.\footnote{We have absorbed the overall coefficient into the definition of ${\cal T}^{\Rc_O}_{1234}$.} Note that, upon decomposing this into a single OPE channel, we must project ${\cal T}$ onto a crossed channel; in doing so, multiple representations will generically appear, not only $\Rc_O$. Such projections were carried out for a specific example where ${\cal G}=SO(8)$ in \cite{Giombi:2017mxl}, involving a double-trace flow from the ABJM theory \cite{Aharony:2008ug}.

\sec{Applications}\label{s5}
In this section, we use our results to derive new double-trace data in interacting vector models in various $d$. We also specialize the operator dimensions to certain values where our results for $\d\g_\ell$ simplify. 

\ssec{$\DP=\D=d-2$: Vector models}
In the special case $\DP=\D=d-2$, our results can be used to extract predictions for the four-point functions and corresponding OPE data 
of certain {\it non-singlet} operators in the $O(n)$ vector model, as we now explain.

Let us start with $n=N M$ free scalars $\varphi^{ia}$, where $i=1,\ldots, N, a=1,\ldots, M$, and take $N$ to be large with $M$ fixed. 
This defines a free CFT with $O(n)$ global symmetry, but we can now look at the spectrum of singlet operators under the $O(N)$ 
symmetry rotating the $i$-index. The single-trace scalar operators of the free CFT in the $O(N)$ 
singlet sector are
\begin{equation}
\Phi^{ab} = \varphi^{i(a}\varphi^{ib)}\,,\qquad O = \varphi^{ia}\varphi^{ia}
\label{singlet-scalars}
\end{equation} 
where $\Phi^{ab}(x)$ is in the symmetric traceless of $O(M)$, and $O(x)$ is a singlet of $O(M)$. Similarly, there are towers of conserved higher-spin 
operators, in the singlet and symmetric traceless of $O(M)$ for even spin, and in the antisymmetric of $O(M)$ for odd spin. 
This singlet sector of the CFT is expected to be dual to 
Vasiliev higher-spin theory in AdS$_{d+1}$ with $O(M)$ Chan-Paton factors \cite{Vasiliev:1999ba}. (See \cite{Giombi:2012ms,Giombi:2016ejx} for reviews of the higher-spin/vector model duality.) 
In particular, the bulk spectrum includes scalar fields dual to the operators in (\ref{singlet-scalars}). 
All of these bulk scalars have the same mass $m^2=-2(d-2)$, and admit two choices of boundary conditions $\Delta=d-2$ or $\Delta=2$. With the former choice, 
the higher-spin theory is dual to the free CFT. Suppose we now impose the alternate $\Delta=2$ boundary condition on the $O(M)$ singlet scalar 
dual to $O= \varphi^{ia}\varphi^{ia}$. This then corresponds to adding the double-trace deformation 
\begin{equation}
\delta S = \l \int d^dx \,(\varphi^{ia}\varphi^{ia})^2 \,,
\label{crit-vec}
\end{equation}
and flowing to the critical vector model in the IR, where $\Delta_O=2+\O(1/N)$. Because we are concentrating on the $O(N)$ singlet sector, we 
can develop the usual $1/N$ expansion, with $\Phi^{ab}$ and $O$ playing the role of ``single-trace'' operators. Then our results can be used to compute the change in the four-point function $\langle \Phi^{a_1 b_1}(x_1) \Phi^{a_2 b_2}(x_2) \Phi^{a_3 b_3}(x_3) \Phi^{a_4 b_4}(x_4) \rangle$ from UV to IR. 

On the other hand, the fixed point of the vector model (\ref{crit-vec}) is just the same as the usual 
critical $O(n)$ model with $n=NM$. From the $O(n)$ point of view, our results give the four-point function of scalar bilinears in non-trivial representations of $O(n)$, and the 
corresponding spinning double-trace anomalous dimensions encoded in it. (Let us stress once again that we would not be able to compute change in the four-point function of $O(x)$ with our result.) 

As a particularly tractable example where we can directly apply our results for pairwise identical operators in Sections \ref{pair-4pt} and \ref{s44}, 
we can consider the case $M=2$. We can then introduce a complex basis 
\begin{equation}
\phi^i = \varphi^{i1}+i \varphi^{i2}\,,\qquad \bar\phi^i = \varphi^{i1}-i \varphi^{i2}\,.
\end{equation}
The two symmetric traceless operators $\Phi^{ab}$ in (\ref{singlet-scalars}) correspond to linear combinations of the complex operators
\begin{equation}
\Phi = \phi^i \phi^i\,,\qquad \Phi^* = \bar\phi^i \bar\phi^i\,,
\end{equation}
with charge $\pm 2$ under $U(1)\simeq SO(2)$, while the singlet is just $O=\phi^i \bar\phi^i$. The change in the four-point function 
\e{}{\la\Phi(x_1)\Phi^*(x_2)\Phi(x_3)\Phi^*(x_4)\ra}
from UV to IR is given by (\ref{pair2}), provided we identify $\Psi=\Phi^*$ and take $\Delta=\Delta_{\Phi}=\Delta_{\Psi}=d-2$.\footnote{Note 
that we have $\la \Phi\Phi^* O\ra \neq 0$ but $\la\Phi\Phi O \ra =0$ due to $U(1)$ charge conservation.} The formula for the change in anomalous dimensions of the double-trace operators $[\P\P^*]_{0,\ell}$ is then given by \eqr{dg0ell} with $\Delta=\Delta_{\Phi}=d-2$, 
and is valid both for even and odd $\ell$. In fact, since the UV theory is free, $\g^{UV}_{n,\ell}=0$. Therefore, in these cases,
\e{}{\d\g_\ell = \g_\ell^{IR}}
and we can use our formulas to read off the anomalous dimensions in the interacting IR CFT. The same observation was made in \cite{Giombi:2017mxl}, where $\g^{UV}_\ell=0$ due to supersymmetry for UV-protected double-trace operators. We will denote $\g^{IR}_{\ell}$ simply by $\g_\ell$ below. For various values of the spacetime dimension $d$ we find, from \eqr{dg0ell},
\es{d36}{d=3:&\quad {\g_{\ell}\o \g_{0}} = {1\o 2\ell+1}\\
d=4:&\quad {\g_{\ell}\o \g_{0}} = {2H_{\ell+1}\o (\ell+1)(\ell+2)}\\
d=5:&\quad {\g_{\ell}\o \g_{0}} = {12\o (\ell+3)(\ell+4)}\\
d=6:&\quad {\g_{\ell}\o \g_{0}} = \frac{18 \left(-8 H_{\ell+3}+\ell (\ell+7)+18\right)}{(\ell+1) (\ell+2) (\ell+5) (\ell+6)}}
where $H_{x}= \sum_{n=1}^x 1/n$ is the harmonic number, and 
\e{auvdm2}{\g_{0} = {8\o N}{\G(d-2)\o \G({d\o2})\G^2({d-2\o2})\G(2-{d\o2})}}
where we used $a^{UV}=4/N$, which can be found by Wick contractions. From (\ref{a00nonidentical}), we can also find the change in OPE coefficient of the double-trace scalar operator, which is simply
\e{}{\d a^{(1)}_{0} = {2\o 2-d} \g_{0}}
Note that the sign is negative for all $d>2$. 

Let us analyze these results. We observe that for all $d>4$, the anomalous dimensions $\gamma_{\ell}$ grow like 
\e{}{\g_{\ell\gg1}\sim \ell^{-2}}
consistent with the lightcone bootstrap \cite{Komargodski:2012ek,Fitzpatrick:2012yx}. This follows from the previous formulas and
\e{}{H_{\ell\gg1} \sim \log\ell+\g_E+O(\ell^{-1})}
where $\g_E$ is the Euler constant. In $d=3$, $\g_{\ell\gg1}\sim\ell^{-1}$; this is also consistent, because the tower of slightly broken higher spin currents with $\t=1+\O(1/N)$ furnishes the leading-twist sector of the $\Phi\times\Phi^*$ OPE instead of $O$. 

The case of $d=4$ is somewhat special: we really should work in $d=4-\eps$, since (to leading order in $1/N$)
\e{phiphistar}{\g_{0}\Big|_{d=4-\eps} \approx {4\o N}\eps+O(\eps^2)~.}
So 
\e{gamell-WF}{\g_{\ell}\Big|_{d=4-\eps} \approx {8\o N}\left({H_{\ell+1}\o (\ell+1)(\ell+2)}\right)\eps+O(\eps^2)}
gives the anomalous dimensions of the $[\P \P^*]_{0,\ell}$ operators at the Wilson-Fisher fixed point of the critical vector model 
(\ref{crit-vec}) in $d=4-\eps$ (for $M=2$ in the present case). At large $\ell$, we see logarithmic behavior,
\e{}{\g_{\ell\gg1}\Big|_{d=4-\eps} \approx {8\o N}{\log\ell\o\ell^2}\eps+\ldots}
We may also write \eqr{gamell-WF} in terms of the conformal spin,\footnote{This follows from the general definition $J^2 \equiv {1\o2}(\D_{\cal O}+\ell)(\D_{\cal O}+\ell-1)$ for the exchange of an operator of dimension $\D_\O$; for us, $\D_\O=2(d-2)+\ell$ to leading order in $1/N$.}
\e{}{J^2 \equiv (\ell+d-2)(\ell+d-3)}
In $d=4-\eps$, 
\e{4deps}{\g_{\ell}\Big|_{d=4-\eps} \approx {8\o N}\left({H_{{1\o 2}(-1+\sqrt{1+4J^2})}\o J^2}\right)\eps+O(\eps^2)}
At large $J$, after the $\log J/J^2$ term, the expansion is in even powers of $J^{-2}$ \cite{Alday:2015eya}. It is interesting to note the similarity of our result to the one obtained in Section 4 of \cite{Alday:2016jfr}, where $\la \phi^2 \phi^2 \phi^2 \phi^2 \ra$ was computed in a small deformation of a free scalar CFT in $d=4$. On general grounds, the first-order anomalous dimensions of the ``single-trace'' currents, $J_\ell = \phi \p^{\ell}\phi$ were found to be $\g(J_\ell) = c_1 + c_2 H_{{1\o 2}(-1+\sqrt{1+4J^2})}$,
where now $J^2 = \ell(\ell+1)$, and the $c_i$ are constants that could not be fixed by symmetries alone. It would be interesting to reproduce our results \eqr{d36} using slightly broken higher spin symmetry, which may give a natural explanation for the appearance of harmonic functions.

The anomalous dimensions (\ref{phiphistar})-(\ref{gamell-WF}) 
may be also computed directly by conformal perturbation theory methods in $d=4-\eps$, at finite $N$; 
as a check of our results, we outline this calculation in Appendix 
\ref{conf-pert} for the case $\ell=0$. The final result is 
\es{finiteN}{
\gamma_{0} = \frac{4 \epsilon}{N+4} {N+1\o N}\,,\qquad \gamma_{\Phi} = \frac{\epsilon}{N+4} }
which in turn can be seen to match our prediction \eqref{phiphistar} at large $N$. Note that in our notation, $\Delta_{\Phi \Phi^*} = \Delta^{(0)}_{\Phi\Phi^*} +4 \gamma_\phi +2\gamma_{\Phi} + \gamma_{0}$ where $\Delta^{(0)}_{\Phi\Phi^*} = 2 (d-2)$ is the classical scaling dimension. 

Recall that, as explained earlier, while in our calculation above we viewed $\P, \P^*$ as single-trace operators in the $O(N)$ singlet sector of a $O(N)\times O(2)$ model, we can view $\P, \P^*$ 
as certain bilinear $O(2N)$ non-singlet operators, which belong to the rank-two symmetric traceless representation of $O(2N)$ 
(this is the only non-singlet representation appearing at the level of scalar bilinears).\footnote{Under the branching $O(2N) \mapsto O(N)\times O(2)$, they are invariant under an $O(N)$ subgroup, but charged under $O(2)$.} 
Hence, our results above can be seen to give the connected four-point function of symmetric traceless scalar bilinear operators in the usual $O(n)$ model,\footnote{Note that, of course, while (\ref{crit-vec}) was written in a ``$O(N)\times O(M)$'' notation, the fixed point has the full $O(n)=O(NM)$ symmetry since the perturbation is a singlet of 
$O(n)$.}  and the anomalous dimensions of their double-trace composites. As far as we know, the result for the anomalous dimensions of the operators $[\Phi \Phi^*]_{0,\ell}$, with $\Phi$ belonging to the rank-two symmetric traceless representation of $O(2N)$, is new.  A closely related result, which was derived by Lang and Ruhl \cite{Lang:1992zw}, are the anomalous dimensions of the singlet double-trace operators $[OO]_{0,\ell}\sim [\sigma \sigma]_{0,\ell}$ (see eq.~(4.49) of \cite{Giombi:2017rhm}). It is interesting to note that in $d=4-\epsilon$, their result has a similar $\sim \log(\ell)/\ell^2$ behavior as our result (\ref{gamell-WF}).   

The calculations of this section can be generalized in a straightforward way to the  $O(N)\times O(M)$ case, where we view $O$ as an $O(M)$ singlet, 
and take $\Phi^{ab}$ in the symmetric traceless representation of $O(M)$. The OPE data encoded in the change of the four-point function 
$\langle \Phi^{a_1 b_1}(x_1) \Phi^{a_2 b_2}(x_2)\Phi^{a_3 b_3}(x_3) \Phi^{a_4 b_4}(x_4) \rangle$
can be extracted introducing $O(M)$ projectors 
as explained in \cite{Giombi:2017mxl}. In fact, since the exchanged operator $O(x)$ 
is an $O(M)$ singlet, the tensor in \eqr{global} is trivial, and role of the projectors 
simply cancels out when expanding in a given OPE channel. Hence, the resulting anomalous dimensions are the same as those listed above in \eqr{d36}.

\ssec{$\DP=d-1, \D=d-2$}
In this case, computing the change in anomalous dimensions of $[\P\P]_{0,\ell}$ operators in the IR, the result \eqr{dg0ell} for low values of $d$ is
\es{d1d2}{d=3:&\quad {\d\g_{\ell}\o\d\g_0} = {2\o 2+\ell}\\
d=4-\eps:&\quad \d\g_{\ell} \approx 8a^{UV}_{\P\P O}\left({H_{\ell+2}-1\o (\ell+1)(\ell+4)}\right)\eps+O(\eps^2)\\
d=5:&\quad {\d\g_{\ell}\o\d\g_0} = {3(40+9\ell)\o (4+\ell)(5+\ell)(6+\ell)}}
where we expanded $\d\g_0$ explicitly near $d=4$ using \eqr{g00}. 

One class of UV CFTs in $d=3$ in this category comes from supergravity compactifications on AdS$_4\times M_7$, in which $M_7$ has non-trivial internal cycles. In particular, for any Sasaki-Einstein $M_7$ with a nonzero second Chern number $b_2$, the CFT has an extra $\N=2$ conserved current multiplet, the Betti multiplet, due to wrapped M2-branes. These multiplets contain $\D=1$ and $\D=2$ scalars that are singlets under all global symmetries, which we identify with $ O$ and $\P$, respectively, in the calculation above. The $\D=1$ Betti scalar is parity odd, so the CFT must break parity in order that $a^{UV}_{\P\P O}\neq 0$. A parity-breaking mechanism using internal fluxes for CFTs with AdS$_4\times M_7$ duals was introduced in the context of the ABJ theory \cite{abj}, where $M_7 = S^7/\Z_k$, and applied to other $M_7$ with $b_2\neq 0$ in e.g. \cite{Martelli:2009ga}. 

Similar simplifications as \eqr{d1d2} occur for other special values of $(\DP,\D)$.

\section*{Acknowledgments}

We thank Igor Klebanov, David Simmons-Duffin, Charlotte Sleight, Massimo Taronna and Herman Verlinde for helpful discussions.  The work of S.G. and V.K. is supported in part by the US NSF under Grant No.~PHY-1620542. E.P. is supported in part by the Department of Energy under Grant No. DE-FG02-91ER40671, and by Simons Foundation grant 488657 (Simons Collaboration on the Nonperturbative Bootstrap). This material is based upon work supported by the U.S. Department of Energy, Office of Science, Office of High Energy Physics, under Award Number DE-SC0011632.

\appendix

\sec{AdS harmonics and propagator identities}\label{appa}

The identity (\ref{conv}) is closely related to the so-called ``split'', or harmonic, representation of the bulk-to-bulk 
propagator (see e.g. \cite{Penedones:2010ue, Fitzpatrick:2011ia, Costa:2014kfa}). Since the bulk-to-bulk propagators with either boundary condition satisfy the same equation (\ref{Green}), their difference must be proportional 
to an AdS harmonic function (see e.g. \cite{Penedones:2007ns} for a review), which may be defined as the solution to
\begin{equation}
\left[\nabla^2_x + \frac{d^2}{4}+\nu^2\right]\Omega_{\nu}(x,y)=0\,,\qquad \,.
\label{harmonic}
\end{equation}
with normalization condition $\int_{-\infty}^{\infty} d\nu \,\Omega_{\nu}(x,y)=\delta^{(d+1)}(x,y)$. 
It is well-known that AdS harmonic functions admit a ``split'' representation as a convolution of bulk-to-boundary propagators (see e.g. 
\cite{Leonhardt:2003qu,Penedones:2010ue,Costa:2014kfa})
\begin{equation}
\Omega_{\nu}(x,y) = \frac{\nu^2}{\pi}\int_{\partial {\rm AdS}} d^d \vec{y}_0 K_{\frac{d}{2}+i\nu}(x;\vec{y}_0)K_{\frac{d}{2}-i\nu}(y;\vec{y}_0)\,.
\label{harmonic-split}
\end{equation}
Noting that we need to take $d^2/4+\nu^2=-m^2=-\Delta(\Delta-d)$, i.e. $\nu=i(\Delta-\frac{d}{2})$, 
and carefully fixing an overall normalization factor (for instance by looking at the coincident point limit),\footnote{The precise proportionality 
constant between the harmonic function and the difference of bulk-to-bulk propagators is found to be  
$$ 
G_{d-\Delta}(x,y)-G_{\Delta}(x,y)= \frac{4\pi}{d-2\Delta} \Omega_{i(\Delta-\frac{d}{2})}(x,y)\,.
$$}
one recovers the identity (\ref{conv}).

Let us also note that a single bulk-to-bulk propagator (as opposed to the difference) can be written, using (\ref{harmonic}) and (\ref{harmonic-split}),  as
\es{G-split}{G_{\Delta}(x,y) &= \int_{-\infty}^{\infty} d\nu \frac{\Omega_{\nu}(x,y)}{\nu^2+(\Delta-\frac{d}{2})^2}\\& = 
 \int_{-\infty}^{\infty} \frac{d\nu\, \nu^2}{\pi(\nu^2+(\Delta-\frac{d}{2})^2)}\int_{\partial {\rm AdS}} d^d \vec{y}_0 K_{\frac{d}{2}+i\nu}(x;\vec{y}_0)K_{\frac{d}{2}-i\nu}(y;\vec{y}_0)}
The formula (\ref{conv}) for the difference of boundary conditions can then be seen to arise from just (twice) the contribution 
of the pole at $\nu=i(\Delta-d/2)$ in the spectral integral above. As an additional remark, note that the split representation (\ref{G-split}), coupled with 
our result for the ``two-triangle'' diagram arising from the difference of boundary conditions, implies that a given exchange diagram with external 
operator $\Phi$ and exchange of a scalar with dimension $\Delta$ can be written as a sum of ${\bar D}$-functions (one for each channel) as in (\ref{4pt-final}), with 
$\Delta$ in the ${\bar D}$-function indices replaced by $d/2+i\nu$, 
and integrated over the spectral parameter $\nu$ with measure determined by (\ref{G-split}).

\sec{Identities for functions $\bar{D}$, $H$ and $G$}
\label{DHG}
Let us recollect here explicit definitions and relations between the functions commonly appearing in the AdS/CFT literature. We follow the notations of \cite{Dolan:2000uw,Dolan:2000ut}.

In AdS/CFT calculations, the $D$-functions are associated to Witten diagrams involving contact interactions \cite{Liu:1998ty,DHoker:1999kzh,Dolan:2000ut}. At the four-point  
level
\begin{equation}
\label{D-func}
D_{\Delta_1\Delta_2\Delta_3\Delta_4}(x_1,x_2,x_3,x_4) = \int \frac{dz \,d^d\vec x}{z^{d+1}} 
\tilde{K}_{\Delta_1}(z,\vec x;x_1) \tilde{K}_{\Delta_2}(z,\vec x;x_2)\ \tilde{K}_{\Delta_3}(z,\vec x;x_3) \tilde{K}_{\Delta_4}(z,\vec x;x_4)\,,
 \end{equation}
where we defined the ``un-normalized'' bulk-to-boundary propagators
\begin{equation}
\tilde{K}_{\Delta_1}(z,\vec x;\vec x') = \left(\frac{z}{z^2+(\vec x-\vec x')^2}\right)^{\Delta}\,.
\end{equation}
The integral in (\ref{D-func}) may be evaluated introducing Schwinger parameters, and yields 
\begin{equation}
D_{\Delta_1\Delta_2\Delta_3\Delta_4}(x_1,x_2,x_3,x_4) = 
\frac{\Gamma\left(\frac{1}{2}\sum_i \Delta_i-\frac{d}{2}\right)}{2\prod_i \Gamma(\Delta_i)}\int_0^{\infty} \prod_i d\alpha_i \alpha_i^{\Delta_i-1} 
\frac{e^{-\frac{1}{\Lambda}\sum_{i<j}\alpha_i \alpha_j x_{ij}^2}}{\Lambda^{\frac{1}{2}\sum_i \Delta_i}}\,, \quad ~~\Lambda \equiv \sum_i \alpha_i
\label{D-Schwinger}
\end{equation}
This can be written in terms of the ``reduced'' ${\bar D}$-functions, which are functions of cross-ratios only, as \cite{Dolan:2000ut}
\begin{equation}
\begin{aligned}
\label{Dbar}
&D_{\Delta_1\Delta_2\Delta_3\Delta_4}(x_1,x_2,x_3,x_4)= 
\frac{\Gamma\left(\Sigma-{d\ov2}\right)}{2\, \prod_i \Gamma\left(\Delta_i\right)}
\frac{x_{14}^{2(\Sigma-\Delta_1-\Delta_4)} x_{34}^{2(\Sigma-\Delta_3-\Delta_4)}}
{x_{13}^{2(\Sigma-\Delta_4)} x_{24}^{2\Delta_2}}\bar{D}_{\Delta_1\Delta_2\Delta_3\Delta_4}(u,v)\\ 
&\Sigma \equiv \frac{1}{2}\sum_i \Delta_i\,.
\end{aligned}
\end{equation}
In the definition above, the powers $\Delta_i$ are arbitrary. In the special case $\Sigma=d/2$, the same $\bar D$-functions arise from the well-known four-point conformal integral
\begin{equation}
\begin{aligned}
I_{\Delta_1\Delta_2\Delta_3\Delta_4}(x_i)&= \int d^d z\frac{1}{(x_1-z)^{2\Delta_1}(x_2-z)^{2\Delta_2}(x_3-z)^{2\Delta_3}(x_4-z)^{2\Delta_4}}\\
&=\frac{\pi^{\frac{d}{2}}}{\prod_i \Gamma(\Delta_i)}\int_0^{\infty} \prod_i d\alpha_i \alpha_i^{\Delta_i-1} 
\frac{e^{-\frac{1}{\Lambda}\sum_{i<j}\alpha_i \alpha_j x_{ij}^2}}{\Lambda^{d/2}}\\
&\stackrel{\sum \Delta_i = d}{=}\frac{\pi^{\frac{d}{2}}}{\Gamma(\Delta_1)\Gamma(\Delta_2)\Gamma(\Delta_3)\Gamma(\Delta_4)} 
\frac{x_{14}^{d-2\Delta_1-2\Delta_4} x_{34}^{d-2\Delta_3-2\Delta_4}}{x_{13}^{d-2\Delta_4} x_{24}^{2\Delta_2}}
\bar{D}_{\Delta_1\Delta_2\Delta_3\Delta_4}(u,v)
\label{Db-conf}
\end{aligned}
\end{equation}
as can be seen by comparing the Schwinger parameter integral in the second line of (\ref{Db-conf}) to (\ref{D-Schwinger}). 

The $\bar D$-functions can be related to the $H$ function defined in \cite{Dolan:2000uw,Dolan:2000ut} as follows:
\e{bard}{\bar{D}_{\D_1 \D_2 \D_3 \D_4} (u,v) = H\left( \D_2, \Sigma -\D_4, \D_1+\D_2 - \Sigma +1 , \D_1+\D_2; u,v \right)}
where $\Sigma = {1\o 2}\sum_i \D_i$ and the function $H$ is given by:
\es{H}{H&(\alpha,\beta,\gamma,\delta;u.v)%= H(\beta, \alpha,\gamma,\delta; u,v) 
={\G(1-\gamma) \o\G(\delta) } \G(\alpha) \G(\beta) \G(\delta - \alpha) \G(\delta - \beta) G(\alpha, \beta,\gamma,\delta;u , 1 -v) \\
&+ u^{1-\gamma}{ \G(\gamma-1)\o \G(\delta-2\gamma+2)} \G(\alpha-\gamma+1) \G(\beta-\gamma+1)  \G(\delta-\gamma+\alpha+1) \G(\delta-\gamma-\beta+1) \\
&\times G(\alpha - \gamma+1, \beta- \gamma+1,2 - \gamma, \delta - 2\gamma+2;u, 1-v).}
The function $G$ may in turn be defined by explicit power series around $u=0$, $v=1$:
\es{G}{G(\alpha,\beta,\gamma,\delta;u,1-v) = \sum_{m,n=0}^\i {(\delta-\alpha)_m (\delta-\beta)_m \o m! (\gamma)_m} {(\alpha)_{m+n} (\beta)_{m+n} \o n! (\delta)_{2m+n} } u^m (1-v)^n}
For (positive) integer $\gamma$ we also get $\log u$ terms in the $H$ function, arising from the gamma functions and Pochhammer symbols in the formulas above,
\e{logH}{H(\alpha,\beta,k,\delta; u, v)|_{\log u}= 
  {(-1)^k \o (k-1)!} { \G(\alpha) \G(\beta) \G(\delta- \alpha) \G(\delta-\beta) \o \G(\delta) } G(\alpha,\beta, k ,\delta;u,1-v)}
These will be required to reproduce the small-$u$ behavior of the sum over the conformal blocks. The power series part of the $H$ function is also modified for integer $\gamma$. We will use following result for $\gamma=1$ \cite{Dolan:2000uw}:
\es{a6}{H(\alpha,\beta,1,\delta;u,v) &= {1\o \G(\delta)}  \G(\alpha)\G(\beta)\G(\delta-\alpha)\G(\delta-\beta) \Bigg(-\log u \, G(\alpha,\beta,1,\delta;u,1-v)\\ &+    \sum_{m,n=0}^\i {(\delta-\alpha)_m (\delta-\beta)_m \o (m!)^2} {(\alpha)_{m+n} (\beta)_{m+n} \o n! (\delta)_{2m+n} } f_{m n} u^m (1-v)^n \Bigg),\\
f_{mn} &\equiv 2 \psi(1+m) + 2 \psi(\delta+2m+n)- \psi(\delta -\alpha+m) -\psi(\delta-\beta+m)\\& - \psi(\alpha+m+n) - \psi(\beta+m+n) }
Also note that the $G$-function obeys a small-$u$ expansion
\e{g2f1app}{G(\a,\b,1,\d;u,1-v) = {}_2F_1(\a,\b,\d;1-v) + O(u)}
The $\Db$-functions obey the following symmetry relations:
\es{Dbsymm}{\Db_{\D_1 \D_2 \D_3 \D_4}(u,v) &= v^{\D_1+\D_4-\Sigma}\Db_{\D_2\D_1\D_4\D_3}(u,v) \\
&= u^{\D_3+\D_4-\Sigma}\Db_{\D_4\D_3\D_2\D_1}(u,v) \\ 
& = \Db_{\D_3 \D_2 \D_1 \D_4}(v,u)\\
&= \Db_{\Sigma - \D_3 \Sigma-\D_4 \Sigma - \D_1 \Sigma-\D_2} (u,v)\\
&=v^{-\D_2} \Db_{\D_1 \D_2 \D_4 \D_3} (u/v, 1/v)\\
& = v^{\D_4 - \Sigma} \Db_{\D_2 \D_1 \D_3 \D_4} (u/v,1/v)}

\sec{Some calculational details}
In this Appendix we collect various odds and ends of the calculations in Sections \ref{s3} and \ref{s4}.

\ssec{$\D+\D_\Psi=\DP-2p$ for $p>0$}
\label{nonzerop}
We can extend the results of Section \ref{cancellation} for all $p$. In general one has to compute the $\Db$-functions in \eqr{pair2} with
\e{}{\D_1 = {d\o 2}-\D-p~, ~~ \D_2 = {d\o 2}+p~, ~~ \D_3 = -p+\eps~, ~~ \D_4 = \D+p~,}
extracting the term of $\O(1/\eps^2)$ in the small $\eps$ limit. Upon doing so we find the following features. First, only the first $p$ terms diverge like $1/\eps^2$. The powers of $u$ range from $u^{\D-{d\o 2}},u^{\D-{d\o 2}+1},\ldots,u^{\D-{d\o2}+p}$. There is never a log term, so the change in anomalous dimensions always vanishes, $\d\g_{n,\ell}=0$. For example, for $p=1$ we find
\es{}{&\Db_{{d\o 2}-\D-1,{d\o 2}+1,-1+\eps,\D+1}(u,v)\approx \frac{\Gamma (\Delta +1) \Gamma \left(\frac{d}{2}-\Delta -1\right)u^{\Delta -{d\o 2}}}{\epsilon ^2}\\&\times\left(\Delta  (d-2 (\Delta +1))+(\Delta +1) (v-1) (d-2 (\Delta +1))-2u (\Delta +1) \right)+O(\eps^{-1})}
and hence
\es{}{\d\F(u,v)\approx -a_{\P\Psi O}^{UV}\Bigg(&u^{\Delta /2} \left(2+\left(\frac{2}{\Delta }+2\right) (v-1)-\frac{4 (\Delta +1) u}{\Delta  (d-2 \Delta -2)}\right) \\+\left({u\o v}\right)^{\D_\P+\D_{\Psi}\o 2}&v^{\Delta /2} \left(2+\left(\frac{2}{\Delta }+2\right) (u-1)-\frac{4 (\Delta +1) v}{\Delta  (d-2 \Delta -2)}\right)\Bigg)}
Note the $d$-independence of the leading term at small $u$, in analogy with the result at $p=0$.

\ssec{Deriving \eqr{dg0ell}}
\label{2F1}
Here we explicitly extract the residue of the contour integral \eqref{gameq}. The relevant part of $\d\F$ is, using \eqr{g2f1},
\es{log-piece}{
\d\F (u,1-x)|_{u^{\Delta_{\Phi}} \log u} &=  \frac{a^{\rm{UV}}_{\Phi \Phi O} \Gamma (\Delta) \Gamma^2 \left(\frac{d-\Delta}{2}\right)}{\Gamma \left(\frac{d}{2}\right) \Gamma ^2\left(\frac{\Delta}{2}\right) \Gamma \left(\frac{d}{2}-\Delta\right)}\\&\times \left( {}_2F_1\left(\frac{d-\Delta}{2},\frac{\Delta}{2},\frac{d}{2},x\right)  +  (1-x)^{\frac{d-\Delta}{2} -\Delta_{\Phi}}  {}_2F_1\left(\frac{d-\Delta}{2},\frac{d-\Delta}{2},\frac{d}{2},x\right) \right)}
From \eqr{gameq}, the residue is given by the coefficient of $x^\ell$ in the product of \eqr{log-piece} with $F_{1-\DP-\ell}(x)$, defined as the hypergeometric function in \eqr{kfunc}. For the first term in \eqr{log-piece} it is rather straightforward to extract the term of order $x^{\ell}$ by multiplying the respective hypergeometric series. For the second term in \eqref{log-piece} one uses the following identity:
\begin{equation}
{}_2 F_1 (a,b,c;x) = (1-x)^{-b} {}_2 F_1 \left(b, c-a, c, \frac{x}{x-1}\right)
\end{equation}
to transform the integrand of \eqr{gameq} to
\begin{equation}
x^{-1-\ell} \left(1-x\right)^{\ell-1} F_{1-\DP-\ell}\left({x\o x-1}\right) {}_2F_1\left(\frac{d-\Delta}{2},\frac{\Delta}{2},\frac{d}{2};{x\o x-1}\right) 
\end{equation}
The final step is to take ${x\o x-1}\rar x$, which transforms the power law prefactor in the integral as
\begin{equation}
dx \,x^{-1-\ell} \left(1-x\right)^{\ell-1} \rar dx\,(-1)^\ell x^{-1-\ell}
\label{sign}
\end{equation}
and only deforms the small contour around $x=0$ without changing the orientation. This proves that both terms contribute the same when $\ell$ is even and cancel out for odd $\ell$. The end result is given in \eqr{gamell}--\eqr{dg0ell}.

\ssec{$\D_\P \neq \D_\Psi$ anomalous dimensions}
\label{generalnonidentical}
Here we give some intermediate steps leading to \eqref{pairdgam}, \eqref{e9a}. To handle the case of pairwise identical operators, we introduce a following generalization of the $F_\beta (x)$  function in \eqref{kfunc}: 
\begin{equation}
F_{\beta; a} (x) =  {}_2F_1 \left( \beta+a, \beta-a, 2\beta , x \right)
\end{equation} 
$x^\beta F_{\beta; a} (x)$ are eigenfunctions of the operator $D_a = x^2(1-x)\p_x^2 - x^2 \p_x +a^2 x$ with eigenvalue $\b(\b-1)$, and obey the orthogonality condition
\e{projpair}{{1\o 2\pi i}\oint_{x=0} x^{\beta-\beta'-1} F_{\beta;a}(x) F_{1-\beta';a}(x) = \delta_{\beta,\beta'}}
with $\b-\b'\in\Z$ and a counterclockwise contour encircling the origin. At $u\ll1$, the conformal blocks take the form:
\e{gcollpair}{g_{\t_n,\ell}(u\ll1,v) \approx g_{n,\ell}^{coll}(v) = x^{\ell}F_{\frac{\D_{\Phi}+\D_\Psi}{2}+n+\ell;{\DP-\D_\Psi \o 2}}(x)}
where $x=1-v$. 

With this in hand we proceed as in Section \ref{s422}. The leading $\log u$ term of $\d\F$ at small $u$ is
\e{logeqnonid}{\d\F^{[\Phi\Psi]}(u,v)\Big|_{\log u} \approx {u^{\frac{\D_{\Phi}+\D_\Psi}{2}}\o 2}\sum_\ell a_{\ell}^{\0}\d\g_{\ell}x^{\ell} F_{\frac{\D_{\Phi}+\D_\Psi}{2}+\ell;{\DP-\D_\Psi \o 2}}(x)}
Applying the orthogonality condition \eqref{projpair} allows us to write down the following generalization of  \eqref{gameq}:
\e{gameqpair}{\d\g_{\ell} = {1\o \pi i a_{\ell}^{\0}}\oint_{x=0}x^{-1-\ell}F_{1-{\D_{\Phi}+\D_\Psi \o 2}-\ell; {\DP-\D_\Psi \o 2}}(x)\, \left[\d\F^{[\Phi\Psi]}(u,1-x)\big|_{u^{{\D_{\Phi}+\D_\Psi \o 2}}\log u}\right]}.
From the explicit form of the four-point function \eqref{pair2}, the relevant piece is
\begin{equation}\label{244}
\begin{aligned}
\d\F(u,v)\Big|_{u^\frac{{\Delta_{\Phi}+\Delta_\Psi}}{2} \log u} & =  {a_{\P\Psi O}^{UV}\G(\D) \G\left({d-\D+\DP-\D_{\Psi}\o 2}\right) \G\left({d-\D-\DP+\D_{\Psi}\o 2}\right) \o \G\left(d \o2 \right)\G\left({d\o 2}-\D\right)   \G\left({\D+\DP-\D_{\Psi}\o 2}\right) \G\left({\D-\DP+\D_{\Psi}\o 2}\right)  }  \\
&\times v^{d-\D - \DP-\D_\Psi \o 2} {}_2 F_{1} \left( \frac{d- \D +\DP -\D_\Psi}{2},\frac{d- \D -\DP +\D_\Psi}{2},\frac{d}{2},1-v\right).
\end{aligned}
\end{equation}
Upon plugging this into \eqref{gameqpair} and extracting the residue, we arrive at \eqref{gamelluneq}, and the final formulas \eqref{pairdgam}, \eqref{e9a}.

\ssec{Adding the $ \P \P \to \Psi \Psi$ channel to \eqr{pair2}}
\label{nonzerochannel}
The result for the change in the four-point function $\la \P\Psi\P\Psi\ra$ in the $ \P \P \to \Psi \Psi$ channel can be written as:
\e{}{\d\F(u,v)\big|_{\P \P \to \Psi \Psi}=-\frac{C_{\P\P O}}{C_{\P\P} \sqrt{C_{OO}}}\frac{C_{\Psi \Psi O}}{C_{\Psi \Psi} \sqrt{C_{OO}}}\frac{\G(\D)}{\G^{4}\left(\frac{\D}{2}\right) \G\left(\frac{d}{2}-\D\right)} u^{\frac{\DP+\D_\Psi}{2}} \Db_{\frac{d-\D}{2},\frac{\D}{2},\frac{d-\D}{2},\frac{\D}{2}}(u,v) }
Let us also define:
\e{}{c^{UV}_{\P \P O} = \frac{C_{\P\P O}}{C_{\P\P} \sqrt{C_{OO}}} \qquad c^{UV}_{\Psi \Psi O} = \frac{C_{\Psi \Psi O}}{C_{\Psi \Psi} \sqrt{C_{OO}}}}
The UV and IR values of these coefficients obey the relation \eqref{aUV-aIR}.
Extracting the anomalous dimensions using the same strategy as before, we may get the following combined result:

\es{twochannels}{\d \gamma_{\ell} &=(-1)^{\ell} {(\D_\Psi)_\ell\o (\DP)_{\ell}}\,  
 {}_4 F_3\Farg{-\ell,\, \ell+\DP+\D_\Psi-1,\,\frac{d- \D -\DP +\D_\Psi}{2},\,\frac{\D-\DP+\D_\Psi}{2}}{\D_\Psi,\,\D_\Psi,\,\frac{d}{2}}{1}   \\
& \times  {2 a_{\P\Psi O}^{UV}\G(\D) \G\left({d-\D+\DP-\D_{\Psi}\o 2}\right) \G\left({d-\D-\DP+\D_{\Psi}\o 2}\right) \o \G\left(d \o2 \right)\G\left({d\o 2}-\D\right)   \G\left({\D+\DP-\D_{\Psi}\o 2}\right) \G\left({\D-\DP+\D_{\Psi}\o 2}\right)} + \\ &+  \frac{2 c^{UV}_{\P \P O} c^{UV}_{\Psi \Psi O}\G(\D)\G^{2}\left({d-\D \o 2}\right)}{ \G\left({d\o 2}\right)  \G\left(\frac{d}{2}-\D\right) \G^{2}\left(\frac{\D}{2}\right)} 
  {}_4 F_3\Farg{-\ell,\, \ell+\DP+\D_\Psi-1,\,\frac{d- \D i}{2},\,\frac{\D}{2}}{\D_\P,\,\D_\Psi,\,\frac{d}{2}}{1} 
}
Notice that when $\DP = \D_\Psi$ we retain the relationship  \eqref{dg0ell} between the $\d \gamma_{\ell}$ and $\d \gamma_{0}$. The terms also cancel each other in that limit for odd $\ell$ and are equal for even $\ell$, as they should. The value of $\d \gamma_{0}$ would have the same dependence on $d$ and $\D$ with the following prefactor:
\e{}{\d \gamma_{0} \sim a_{\P\Psi O}^{UV}+c^{UV}_{\P \P O} c^{UV}_{\Psi \Psi O} \xrightarrow[]{\P \equiv \Psi} a_{\P \P O}^{UV}}
as it should, since the three-point coefficient would have two channels contributing in this case.

We now write down the result of the large spin expansion. It fits the general predictions of the lightcone bootstrap \cite{Komargodski:2012ek,Fitzpatrick:2012yx}:
\es{twochannelsinf}{ \d \gamma_{\ell}  \approx & {2 a^{UV}_{\P \Psi O} \G(\D) \G(\DP) \G(\D_\Psi)\o \G\left( {\D+\DP - \D_\Psi \o 2 }\right)  \G\left( {\D-\DP +\D_\Psi \o 2 }\right) \G^{2}\left( {\DP + \D_\Psi \o 2} - {\D \o 2}\right)} {(-1)^{\ell}\o \ell^{\D}} \\- & {2 a^{IR}_{\P \Psi O} \G(d-\D) \G(\DP) \G(\D_\Psi)\o \G\left( {d-\D+\DP - \D_\Psi \o 2 }\right)  \G\left( {d-\D-\DP +\D_\Psi \o 2 }\right) \G^{2}\left( {\DP + \D_\Psi \o 2} - {d-\D \o 2}\right)} {(-1)^{\ell}\o \ell^{d-\D}} \\ + & {2 c^{UV}_{\P \P O} c^{UV}_{\Psi \Psi O}  \G(\D) \G(\DP) \G(\D_\Psi) \o \G^{2}\left( {\D \o 2}\right) \G\left(\DP-{\D\o 2}\right) \G\left( \D_\Psi-{\D\o 2} \right) }  {1\o \ell^\D}  -  {2 c^{IR}_{\P \P O} c^{IR}_{\Psi \Psi O}  \G(d-\D) \G(\DP) \G(\D_\Psi) \o \G^{2}\left( {d-\D \o 2}\right) \G\left(\DP-{d-\D\o 2}\right) \G\left( \D_\Psi-{d-\D\o 2} \right) }  {1\o \ell^{d-\D}}} 

We may also write down the result for the $\d a_{0}^{(1)}$ coefficient:
\es{a00nonidentical}{\d a_{0}^{(1)} &=  { a_{\P\Psi O}^{UV}\G(\D) \G\left({d-\D+\DP-\D_{\Psi}\o 2}\right) \G\left({d-\D-\DP+\D_{\Psi}\o 2}\right) \o \G\left(d \o2 \right)\G\left({d\o 2}-\D\right)   \G\left({\D+\DP-\D_{\Psi}\o 2}\right) \G\left({\D-\DP+\D_{\Psi}\o 2}\right)}   \Biggl(  \psi\left( {\D +\DP-\D_\Psi \o 2}\right)+ \psi\left({\D -\DP+\D_\Psi \o 2} \right) \\& +  \psi\left( {d-\D +\DP-\D_\Psi \o 2} \right) +  \psi\left( {d-\D -\DP+\D_\Psi \o 2 }\right)- 2 \psi\left( {d \o2}\right) +2 \gamma   \Biggr) \\ &+    \frac{2 c^{UV}_{\P \P O} c^{UV}_{\Psi \Psi O}\G(\D)\G^{2}\left({d-\D \o 2}\right)}{ \G\left({d\o 2}\right)  \G\left(\frac{d}{2}-\D\right) \G^{2}\left(\frac{\D}{2}\right) } \Biggl( \psi\left({\D \o 2} \right) + \psi\left(d-\D \o 2\right) -\psi\left({d \o 2}\right) +\gamma \Biggr) }

\sec{Conformal Perturbation Theory for the ``$O(N)\times O(2)$'' model in $d=4-\epsilon$}
\label{conf-pert}
In this Appendix we briefly review the framework of conformal perturbation theory to calculate the anomalous dimensions of the $\P$ and $\P \P^*$ fields in the $O(N)\times O(2)$ model in $d= 4 - \epsilon$. We may view the action as a sum of free theory of $2 N$ scalar fields and a perturbation by an operator $O^2$:
\e{epsaction}{S = S_{0} + \lambda \int d^d x ~O^2(x)}
where $O(x) = \bar{\phi}^i \phi^i$ is the singlet operator with tree-level dimension $\D_0 = d-2$, and $S_0=\int d^d x \partial_{\mu}\phi^i \partial^{\mu}\bar\phi^i$. 

We will need to first study the two-point function of $\Phi = \phi^i \phi^i$ and $\Phi^* = \bar{\phi}^i \bar{\phi}^i$ to the first order in perturbation theory:
\es{d2}{\langle \P(x) \P^*(y) \rangle_{\lambda} &= \langle \P(x) \P^*(y) \rangle_{0} - \lambda \int d^ d z \langle \P(x) \P^*(y) O^2(z)  \rangle_{0} \\ &= {C_2 \o |x-y|^{2\D_{0}}} - \lambda C_3 \int d^d z {1\o |y-z|^{2\D_0} |x-z|^{2\D_0}},}
where $C_2$ and $C_3$ are two- and three-point function coefficients respectively in the unperturbed theory:
\es{}{\langle \P(x) \P^*(y) \rangle_{0} &= {C_2 \o |x-y|^{2\D_0}}\\ \langle \P(x) \P^*(y) O^2(z) \rangle_{0} &= {C_3 \o  |x-z|^{2 \D_0} |y-z|^{2 \D_0}}}
The values of $C_2$ and $C_3$ are obtained by direct Wick contractions in the unperturbed theory: 
\es{}{C_2 = 2 N C^2_{\phi \bar{\phi}} \\ C_3 = 4 N C^4_{\phi \bar{\phi}}}
where $C_{\phi \bar{\phi}}$ is defined as:
\e{}{ \langle \phi^i(x) \bar{\phi}^j(y)  \rangle =  {C_{\phi \bar{\phi}}\delta^{ij} \o  |x-y|^{2\D_\phi}} ,\qquad   C_{\phi \bar{\phi}} = {1\o 2 \pi^2}}
where the $2$ in the denominator follows from our definition of the $\phi$ field: $\phi^i = \varphi^{i,1}+i \varphi^{i,2}$ where $\varphi^{i,a}$, $i=1,\ldots, N$, $a=1,2$ are real scalar fields with canonical normalization.

 Using the integral
\e{}{\int d^d z {1\o |x-z|^{2 \alpha}  |y-z|^{2 \beta} } = {\pi^{{d\o2}}\o |x-y|^{2 (\alpha +\beta - {d\o2})}} {\G\left({d\o2} -\alpha \right) \G\left({d\o2} -\beta \right) \G\left(\alpha+ \beta - {d\o2}  \right)\o \G(\alpha) \G(\beta) \G(d-\alpha-\beta)}} 
we find, after plugging $d=4-\epsilon$ and $\D_0 = 2 -\epsilon$ into \eqr{d2} to the leading order in $\epsilon$,
\es{}{\langle \P(x) \P^*(y) \rangle_{\lambda} &= {C_2 \o |x-y|^{2\D_{0}}} - \lambda {C_3\o |x-y|^{2\D_0}} {4 \pi^2 |x-y|^{\epsilon}\o \epsilon} \\&= {C_2 \o |x-y|^{2\D_{0}}} - \lambda {C_3\o |x-y|^{2\D_0}} 4 \pi^2 \left({1\o \epsilon} + \log |x-y| + O(\epsilon) \right)}
We may now exploit the fact that we are only interested in the anomalous dimension at the conformal fixed point of the $O(2N)$ model, $\lambda^* = {\pi^2 \epsilon \o N+4}$  \cite{Wilson:1972cf}, such that the full two-point function in the interacting theory has to be power-law:
\es{}{\langle \P(x) \P^*(y) \rangle_{\lambda} = {\mathcal{C}_2(\lambda) \o |x-y|^{2(\D_{0} + \gamma(\lambda))}} = {C_2 \o |x-y|^{2 \D_0}  } - {C_2\o |x-y|^{2 \D_o}} 2\gamma \log |x-y| + ...}
where $\gamma(\lambda) = \D - \D_0 = 2\gamma_\phi + \gamma_\P$ and $...$ stands for terms which either do not have $\log$s  or are higher order in $\lambda$. Then we may identify the logarithmic terms to find $\gamma(\lambda)$ to lowest order:\footnote{A more accurate argument introducing the multiplicative renormalization of $\P$ would lead to the same result.}
\e{}{\gamma_{\P} = {2 \pi^2 C_3 \o C_2} \lambda^* = {\epsilon\o N+4}}
Notice that $\gamma_\phi$ is of order $\epsilon^2$, as well known, and doesn't contribute at lowest order. Then the final result can be matched to the well-known anomalous dimension of the symmetric traceless tensor of the $O(2 N)$ model \cite{Wilson:1972cf,Rychkov:2015naa}. 

A similar calculation for the $\P \P^*$ operator may be carried out repeating the same steps as above, where $\tilde{C}_2$ and $\tilde{C}_3$ are now defined through:
\es{}{\langle \P\P^*(x) \P\P^*(y) \rangle_{0} &= {\tilde{C}_2 \o |x-y|^{4\D_0}}\\ \langle \P\P^*(x) \P\P^*(y) O^2(z) \rangle_{0} &= {\tilde{C}_3 \o |x-y|^{2\D_0} |x-z|^{2 \D_0} |y-z|^{2 \D_0}}.}
The $z$-dependence of the three-point function is the same as in the previous calculation, and so will be the value of the integral. To extract $\gamma_{\P \P^*}$ we also recall the definition of the anomalous dimension for a composite operator:
\e{}{\D_{\P \P^*} = 2 \D_{0} + 4\gamma_\phi + 2\gamma_\P + \gamma_{\P \P^*}. }
Here again, $\gamma_\phi$ will not matter, being of order $O(\epsilon^2)$. Expanding the exact two-point function and calculating $\tilde{C}_2$ and $\tilde{C}_3$, we get:
\es{}{ \tilde{C}_2 & =  4N^2 C^4_{\phi \bar{\phi}} \\  \tilde{C}_3 &=  16N(3N+2)  C^6_{\phi \bar{\phi}} } 
and 
\e{}{\gamma_{\P\P^*} + 2\gamma_\P =  {2 \pi^2 \tilde{C}_3 \o \tilde{C}_2} \lambda^* = {2 \epsilon \o N+4} {3N+2 \o N}} 
from which we get 
\e{}{\gamma_{\Phi \Phi^*} = \frac{4 \epsilon}{N+4} {N+1\o N}}
in agreement with \eqref{finiteN}.

\bibliographystyle{ssg}
\bibliography{4pt}

\end{document}